\documentclass[floatfix,twocolumn,showpacs,prl,amsmath,amssymb,preprintnumbers,superscriptaddress]{revtex4-2}
\usepackage{amsmath,amssymb,amsfonts}
\usepackage{bm}
\usepackage{siunitx}
\usepackage{txfonts}

\usepackage[colorlinks=true, bookmarks=true,
bookmarksnumbered=true, bookmarkstype=toc, linkcolor=blue,
urlcolor=blue, citecolor=blue]{hyperref}

\usepackage{mathrsfs} 
\usepackage{braket}  

\newcommand{\soeji}[1]{\mbox{\scriptsize{#1}}}   
\newcommand{\Tc}[0]{T_{\mathrm c}}

\usepackage{graphicx} 
\usepackage{color} 
\usepackage{here}
\usepackage[FIGTOPCAP,nooneline]{subfigure}

\usepackage{ascmac}

\begin{document}

\title{Current-induced superconducting anisotropy of {\boldmath ${\rm Sr_2 Ru O_4}$}}
\author{R.~Araki}
\email{araki.ryo.43w@st.kyoto-u.ac.jp}
\affiliation{Department of Physics, Kyoto University, Kyoto 606-8502, Japan}
\author{T.~Miyoshi}
\affiliation{Department of Physics, Kyoto University, Kyoto 606-8502, Japan}
\author{H.~Suwa}
\affiliation{Department of Physics, Kyoto University, Kyoto 606-8502, Japan}
\author{E.~I.~Paredes Aulestia}
\affiliation{Department of Physics, The Chinese University of Hong Kong, Shatin N.T., Hong Kong}
\author{K.~Y.~Yip}
\affiliation{Department of Physics, The Chinese University of Hong Kong, Shatin N.T., Hong Kong}
\author{Kwing~To~Lai}
\affiliation{Department of Physics, The Chinese University of Hong Kong, Shatin N.T., Hong Kong}
\author{S.~K.~Goh}
\affiliation{Department of Physics, The Chinese University of Hong Kong, Shatin N.T., Hong Kong}
\author{Y.~Maeno}
\affiliation{Department of Physics, Kyoto University, Kyoto 606-8502, Japan}
\author{S.~Yonezawa}
\email{yonezawa.shingo.3m@kyoto-u.ac.jp}
\affiliation{Department of Physics, Kyoto University, Kyoto 606-8502, Japan}

\date{\today}

\begin{abstract}
    In the unconventional superconductor ${\rm Sr_2 Ru O_4}$, unusual first-order superconducting transition has been observed
     in the low-temperature and high-field region, accompanied by a four-fold anisotropy of the in-plane upper critical
     magnetic field $H_{\soeji{c2}}$.
    The origin of such unusual $H_{\soeji{c2}}$ behavior should be closely linked to the debated superconducting symmetry of this oxide.
    Here, toward clarification of the unusual $H_{\soeji{c2}}$ behavior, we performed the resistivity measurements
     capable of switching in-plane current directions as well as precisely controlling the field directions.
    Our results reveal that resistive $H_{\soeji{c2}}$ under the in-plane current
     exhibits an additional two-fold anisotropy. 
    By systematically analyzing $H_{\soeji{c2}}$ data taken under various current directions,
     we succeeded in separating the two-fold $H_{\soeji{c2}}$ component into the one originating from 
     applied current and the other originating from certain imperfection in the sample.
    The former component, attributable to vortex flow effect, is weakened at low temperatures
     where $H_{\soeji{c2}}$ is substantially suppressed. 
    The latter component is enhanced in the first order transition region, possibly reflecting a change in
     the nature of the superconducting state under high magnetic field.
\end{abstract}

\maketitle

\section{Introduction}

The layered perovskite superconductor ${\rm Sr_2 Ru O_4}$ with the transition temperature $T_{\soeji{c}}$ of 1.5 K~\cite{Maeno1994}
 has been extensively studied due to its unconventional pairing state.
Its Fermi surface has a relatively simple topology, consisting of
 three cylindrical sheets ($\alpha$, $\beta$, and $\gamma$)~\cite{Bergemann2003.AdvPhys.52.639,Damascelli2000.PhysRevLett.85.5194}
 with its well-characterized Fermi-liquid behavior.
Recent nuclear magnetic resonance (NMR) experiments using low rf pulses revealed
 a technical problem in previous reports and clarified a reduction of the $^{17}$O Knight shift
 in the superconducting states~\cite{Pustogow2019.Nature.574.72,Ishida2020.JPSJ.89.034712,Chronister2021.PNAS.118.25}.
Recent polarized neutron scattering experiments performed under lower fields also revealed the reduction in the magnetic susceptibility
 in the Ru site~\cite{Petsch2020.Phys.Rev.Lett.125.217004}.
These results cannot be explained in the framework of the traditional spin-triplet superconductivity scenario.
Zero-field muon spin rotation ($\mu$SR) as well as magneto-optic Kerr effect experiments showed evidence for
 time-reversal-symmetry breaking (TRSB)~\cite{Luke1998.Nature.394.558,Xia2006.PhysRevLett.97.167002}.
Recent $\mu$SR experiments under uniaxial stress reveal stress-induced splitting between the onset temperatures
 of superconductivity and TRSB~\cite{Grinenko2021.Nat.Phys.17.748}
 whereas such splitting does not occur as long as the tetragonal symmetry is preserved \cite{Grinenko2021NatureComm.12.3920}.
Furthermore, ultrasound measurements show that the superconducting order parameter of ${\rm Sr_2 Ru O_4}$ consists
 of at least two components~\cite{Benhabib2021.Nat.Phys.17.194,Ghosh2021.Nat.Phys.17.199,Grinenko2021.Nat.Phys.17.748}.
More recently, it is revealed that the NMR spectrum near the upper critical magnetic field $H_{\soeji{c2}}$ exhibits
 characteristic ``double-horn" structure, indicating the superconducting spin smecticity~\cite{Ichioka2007.PhysRevB.76.014503}. 
This result suggests that ${\rm Sr_2 Ru O_4}$ is a strong candidate for the formation of
 the Fulde-Ferrell-Larkin-Ovchinnikov (FFLO) state
~\cite{Fulde1964PhysRev.135.A550,Larkin1964ZhEksp}.

Toward clarification of the debated superconducting order parameter of ${\rm Sr_2RuO_4}$, 
 properties near $H_{\soeji{c2}}$ is of primary importance.
Indeed, the superconducting transition becomes the first-order transition (FOT) under magnetic fields aligned accurately
 in the $ab$ plane and below 0.8~K~\cite{Yonezawa2013.PhysRevLett.110.077003,Yonezawa2014.JPhysSocJpn.83.083706}.
Considering the recent revival of the NMR data, this first-order transition can be well interpreted as a consequence of
 the Pauli-paramagnetic pair-breaking effect, which is not allowed in the traditional chiral-$p$-wave spin-triplet scenario.
More interestingly, in the same temperature region,
 $H_{\soeji{c2//{\it ab}}}$ exhibits a clear four-fold in-plane anisotropy
 \cite{Yonezawa2014.JPhysSocJpn.83.083706,Mao2000.PhysRevLett.84.991,Kittaka2009.PhysRevB.80.174514}. 
Although this four-fold anisotropy preserves the tetragonal crystalline symmetry,
 its origin and relation to the first-order transition have not been clarified.
The motivation of our study is to reveal the relationship between the $H_{\soeji{c2//{\it ab}}}$ anisotropy
 and FOT.

We focus on the anisotropy of $H_{\soeji{c2}}$ in the $ab$ plane under in-plane currents.
First of all, $H_{\soeji{c2}}$ under currents is expected to show the same four-fold symmetry as that under zero current
~\cite{Yonezawa2014.JPhysSocJpn.83.083706,Mao2000.PhysRevLett.84.991,Kittaka2009.PhysRevB.80.174514}.
Secondly, since type-II superconductors under currents and magnetic fields above the lower critical 
 field $H_{\soeji{c1}}$ exhibit
 vortex flow resistivity caused by the Lorentz force~\cite{MatsushitaText},
 $H_{\soeji{c2}}$ in the $ab$ plane of ${\rm Sr_2RuO_4}$ is
 also expected to exhibit current-induced two-fold symmetry depending on the relative angle $\phi_{H}-\phi_I$
 between the magnetic field $H$ and current $I$.
Here, $\phi_H$ and $\phi_I$ are the in-plane angles of the field and current 
 with respect to the crystal axes, respectively.
Thirdly, $H_{\soeji{c2}}$ under currents may also depend on $\phi_I$.
Furthermore, these anisotropies of $H_{\soeji{c2}}$ under currents are expected to show different behavior 
 between the FOT and second-order transition (SOT) regions due to the difference in the dominant pair-breaking mechanisms.
Thus, we consider the direction and strength of the current as new parameters of controlling $H_{\soeji{c2}}$,
 and extensively investigate the temperature dependence of the anisotropies of $H_{\soeji{c2}}$.

 \begin{figure*}[htb]
  \begin{center}
    \def\subfigcapskip{-3pt}
   \subfigure[$\qquad \qquad \qquad \qquad \qquad \qquad \qquad \qquad \quad$]{
    \includegraphics[width=.65\columnwidth]{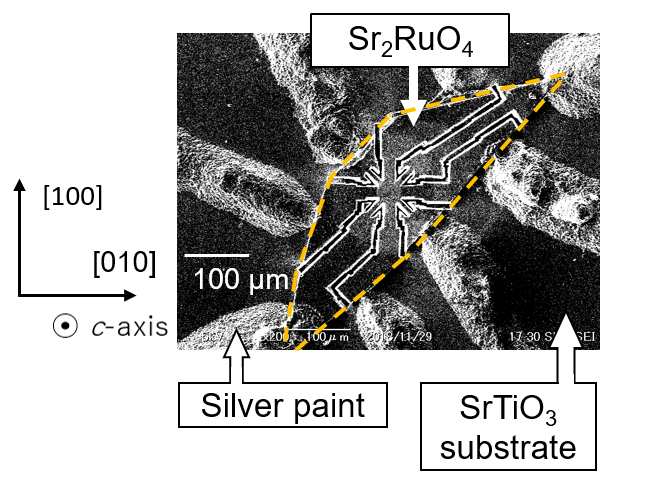} 
   }
   \def\subfigcapskip{-3pt}
   \subfigure[$\qquad \qquad \qquad \qquad \qquad \qquad \qquad$]{
    \includegraphics[width=.65\columnwidth]{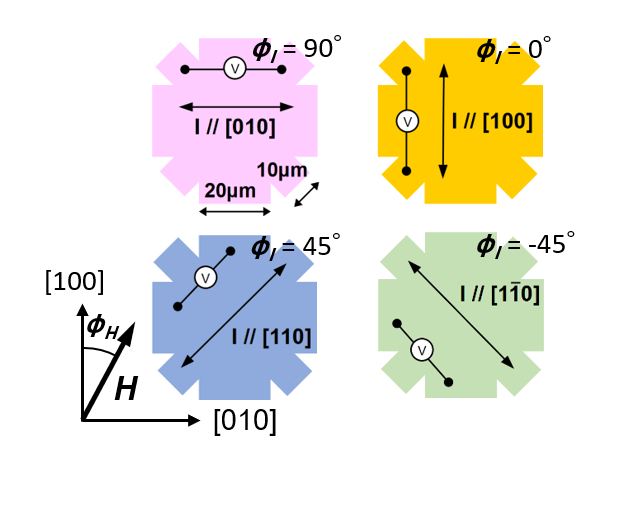}
   }
   \def\subfigcapskip{-3pt}
   \subfigure[$\qquad \qquad \qquad \qquad \qquad \qquad \qquad \qquad \qquad$]{
    \includegraphics[width=.65\columnwidth]{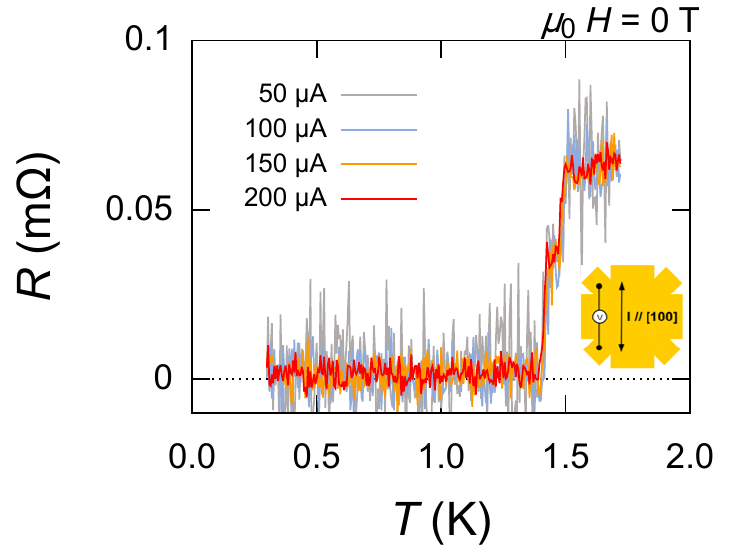} 
   }
  \end{center}
  \caption{Current-direction switchable device of single-crystalline ${\rm Sr_2 Ru O_4}$
  for $H_{\soeji{c2}}$-anisotropy measurements. 
  (a) Scanning electron microscope (SEM) image of the device after
   the focused ion-beam (FIB) process.
  FIB-cut trenches are visible as black lines with white edges.
  The orange broken line indicates the original shape of the ${\rm Sr_2 Ru O_4}$ crystal.
  To obtain low contact resistance, we used high-temperature-cure silver paint (Dupont, 6838).
  (b) Schematics of the four configurations of resistivity measurements
   with different current directions under magnetic fields.
  The [100] direction is defined as $\phi = 0^{\circ}$ and the [010] direction as $\phi = 90^{\circ}$
  for both the field angle $\phi_H$ and current angle $\phi_I$.
  (c) Resistive transition of the device measured on warming under zero field.
  The superconducting transition temperature $T_{\soeji{c}}$ (1.45 K) remains close to
   the value for the cleanest samples (1.5 K)~\cite{Mackenzie1998.PhysRevLett.80.161} even after the FIB process,
   and $\Tc$ does not depend on current below 200~$\muup$A.
 }
\label{Fig:device} 
\end{figure*}

In order to clarify the electric current effect on the anisotropy of $H_{\soeji{c2}}$,
 it is crucial to identify and eliminate the effects of technical origins.
First, because the value of $H_{\soeji{c2//{\it c}}}$ is about one twentieth less than that of $H_{\soeji{c2//{\it ab}}}$~\cite{Kittaka2009.PhysRevB.80.174514},
  the presence of a small misalignment of in-plane magnetic field itself can lead to an apparent two-fold anisotropy of $H_{\soeji{c2}}$ in the $ab$ plane.
Second, since the distribution of currents depends on the shape and
  distortion of the sample device, they can cause apparent anisotropy dependent on the direction of the current $\phi_{I}$.
 To avoid detecting these extrinsic anisotropies,
  we established a measurement procedure in which
  magnetic fields are applied in the $ab$ plane accurately and precisely, and currents can be switched in various crystal orientations.

In this work, we measured resistivity of a micro-structured single-crystalline ${\rm Sr_2 Ru O_4}$ sample
 under various in-plane field and current directions.
The sample was processed by focused ion-beam (FIB) as shown in Fig.~\ref{Fig:device}(a);
 this structured sample allows us to switch the current in the $[100], [010], [110]$ and $[1 \overline{1} 0]$ directions.
We find that, with a given current direction, $H_{\soeji{c2}}$ exhibits two-fold anisotropy as a function of the in-plane field angle,
 in addition to the ordinary four-fold anisotropy.
Careful analysis reveals that the two-fold component consists of
 two contributions: the one caused by external current likely due to vortex-flow effect,
 and the other originating from sample inhomogeneity probably introduced by the FIB process.
We revealed that the latter is enhanced below 0.8 K, namely in the FOT regime, attributable
 to a change in the superconducting order parameter.

\section{Experimental}

In this study, we used ${\rm Sr_2 Ru O_4}$ single crystals grown with the floating-zone method \cite{Mao2000.MaterResBull.35.1813}.
Before the micro-fabrication of the resistivity device, $\Tc$ of a thin crystal (batch: C432; 500~$\muup$m $\times$
 100~$\muup$m $\times$ 10~$\muup$m) was measured to be 1.43 K (see Supplemental Material (SM) \cite{SupplementalAraki}),
 defined as the peak of the imaginary part of the AC susceptibility
 measured using a compact susceptometer~\cite{Yonezawa2015.RevSciInstrum.86.093903}
 that fits inside a commerical refrigerator (Quantum Design; PPMS adiabatic demagnetization refrigerator option).
The crystalline orientation of the sample was determined by x-ray Laue pictures.
The crystal was placed on a single-crystalline ${\rm Sr Ti O_3}$ substrate (Shinkosha), which has a thermal contraction matching 
 with that of ${\rm Sr_2 Ru O_4}$ \cite{Loetzsch2010APL.96.071901,Chmaissem1998PhysRevB.57.5067}.
 After the electrical contact between the ${\rm Sr_2 Ru O_4}$ crystal and eight electrodes, made of high-temperature-cure silver paint
 (Dupont, 6838), were established similarly to previous studies~\cite{Yasui2017PhysRevB.96.180507},
 the surface of the crystal was protected by evaporating a 0.5-$\mu$m layer of SiO$_{x}$.
We then used a focused ion beam (FIB) instrument (JEOL, JIM-4501) with a Ga ion beam
 to fabricate the current-direction switchable resistivity device.
Figure~\ref{Fig:device}(a) shows a scanning electron microscope (SEM) image of the device
 taken from the $c$-axis direction.
 As shown in the figure, the widths of the arms along the [100] and [010] directions are
 both $20~\muup$m,
 and those along the [110] and $[1\overline{1}0]$ directions are $10~\muup$m.
 The device thickness is $10~\muup$m.


The ${\rm Sr_2 Ru O_4}$ device was cooled down to 0.12 K with a ${}^3$He-${}^4$He
 dilution refrigerator (Oxford Instruments, Kelvinox 25).
For controlling the field orientation at low temperatures, we used a vector magnet system
 (Cryomagnetics, VSC-3050) with orthogonally arranged SC magnets
 that generate a horizontal field $H_r$ of up to 5~T
 and a vertical field $H_z$ of up to 3~T to control the polar field angle 
 $\theta_{\soeji{lab}} = \arctan (H_z/H_r)$~\cite{Deguchi2004RSI}.
This vector magnet allows control of $\theta_{\soeji{lab}}$
 with a typical precision of $\Delta \theta_{\soeji{lab}} = 0.005^{\circ}$ at $\mu_{\soeji{0}}H = 1$~T.
Moreover, the magnet set is placed on a horizontal rotation stage, which can control the azimuthal field
 angle $\phi_{\soeji{lab}}$ with a precision of $\Delta\phi_{\soeji{lab}} = 0.001^{\circ}$.
The quasi-two-dimensional (2D) anisotropy of $H_{\soeji{c2}}$ of ${\rm Sr_2 Ru O_4}$ ($H_{\soeji{c2//{\it ab}}}/H_{\soeji{c2//{\it c}}}\sim 20$~\cite{Kittaka2009.PhysRevB.80.174514})
 allows us to align the field to the $ab$ plane
 accurately and precisely: by rotating magnetic field close to $H_{\soeji{c2//{\it ab}}}$,
 we can find a sharp drop of resistivity when the field is exactly parallel to the $ab$ plane (see SM \cite{SupplementalAraki}).
The in-plane alignment was done based on the known four-fold anisotropy of the in-plane $H_{\soeji{c2}}$
 (i.e. $H_{\soeji{c2//[110]}} > H_{\soeji{c2//[100]}}$ below 0.8~K)~\cite{Mao2000.PhysRevLett.84.991,Kittaka2009.PhysRevB.80.174514}.
All the data shown in this paper are plotted using the field angles $\phi_H$ and $\theta_H$ defined in the sample coordinate.


To measure the sample resistance $R$, we employed a DC method with current 
 sign reversal to avoid influence of voltage offsets such as thermoelectric voltages.
We used a combination of a current source (Keithley, 6221) and a nanovoltometer (Keithley, 2182A).
We measured $R$ against various parameters: 
 temperature $T$, magnitude and azimuthal angle of magnetic field $H$ and $\phi_H$,
 current amplitude and direction $I$ and $\phi_I$. 
Both $\phi_H$ and $\phi_I$ are defined with respect to the [100] direction as shown in Fig.~\ref{Fig:device}(b).
We note that for a given current value, 
 the current density for an orthogonal measurement ($I$//[100] or [010])
 is half of that for a diagonal measurement ($I$//[110] or $[1\overline{1}0]$)
 due to the larger width of the arm along the [100] or [010] directions.

Figure~\ref{Fig:device}(c) shows the temperature dependence of $R$ 
 under zero field. The resistivity reaches zero at $\Tc =1.45$~K.
The consistent $\Tc$ values before and after FIB assure that a possible damage or strain due to FIB is
 minimal.
In this study we used the current values below 200~$\muup$A for which $\Tc$ does not change under zero field.

\section{Results and Analysis}

\begin{figure} 
    \centering
    \includegraphics[keepaspectratio, scale=0.90]{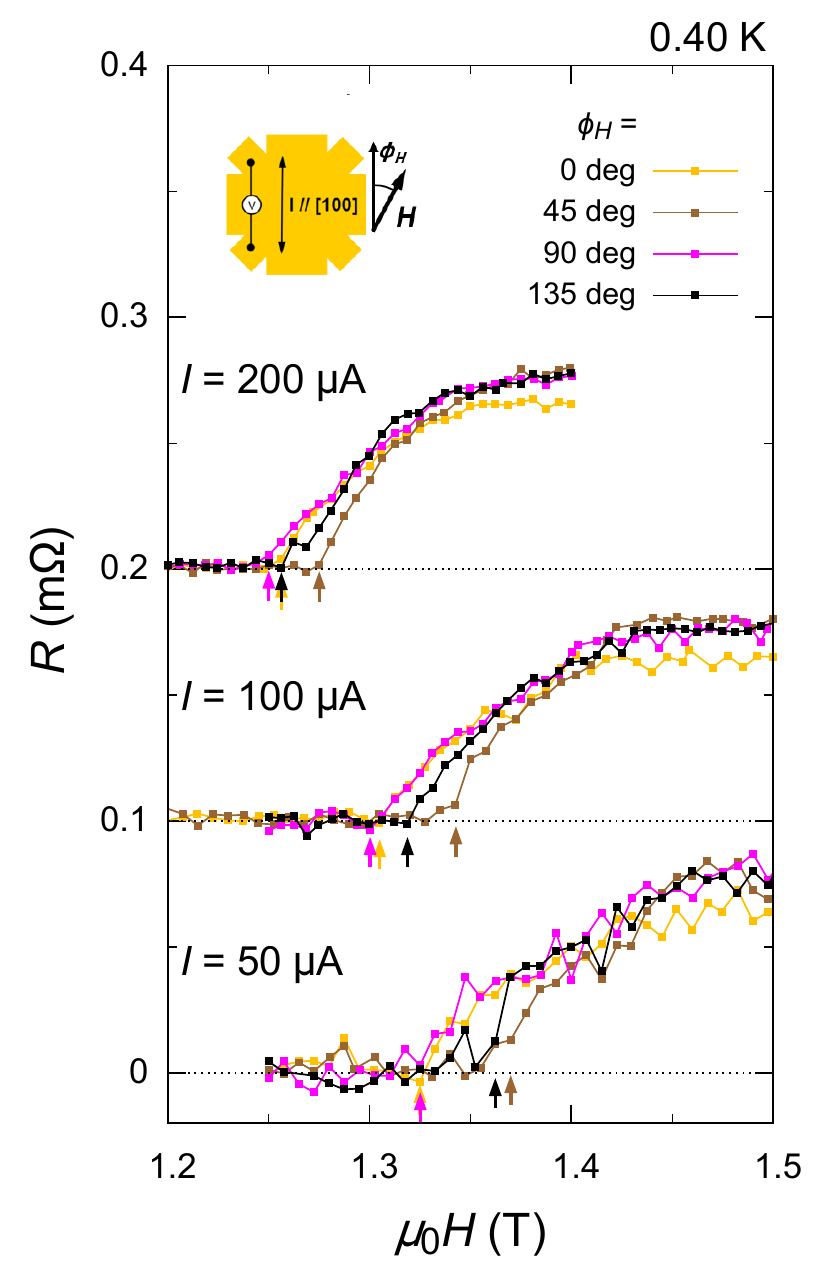}
  \caption{Representative resistive transition of our ${\rm Sr_2 Ru O_4}$ device under various values
   of in-plane currents $I$.
   These data were taken at 0.40 K with current along the [100] direction.
   The curves with $I \geq 100~\muup$A have vertical offsets.
   The yellow, brown, pink and black data points are obtained under $\phi_H = 0^{\circ}, 45^{\circ}, 90^{\circ} $
    and $ 135^{\circ}$, respectively.
   In addition to the known four-fold anisotropy, $H_{\soeji{c2}}$, indicated by the vertical arrows,
    exhibits two-fold anisotropy as evidenced by the differences in the curves 
    between $\phi_H = 0^{\circ}$ and $90^{\circ}$, or between $\phi_H = 45^{\circ}$
    and $135^{\circ}$.
          } \label{082}
\end{figure}

To investigate the anisotropy of the resistive $H_{\soeji{c2}}$ in the $ab$ plane under in-plane currents,
we measured $R(H)$ under various field and current conditions.
We show in Fig.~\ref{082} representative field dependence of the resistance for $I//[100]$.
Additional raw data are shown in SM \cite{SupplementalAraki}.
The in-plane $H_{\soeji{c2}}$, defined with the deviation from $R=0$ as indicated by the arrows, clearly
 depends on the in-plane field direction and on the current strength.


\begin{figure} 
  \centering
  \includegraphics[keepaspectratio, scale=0.88]{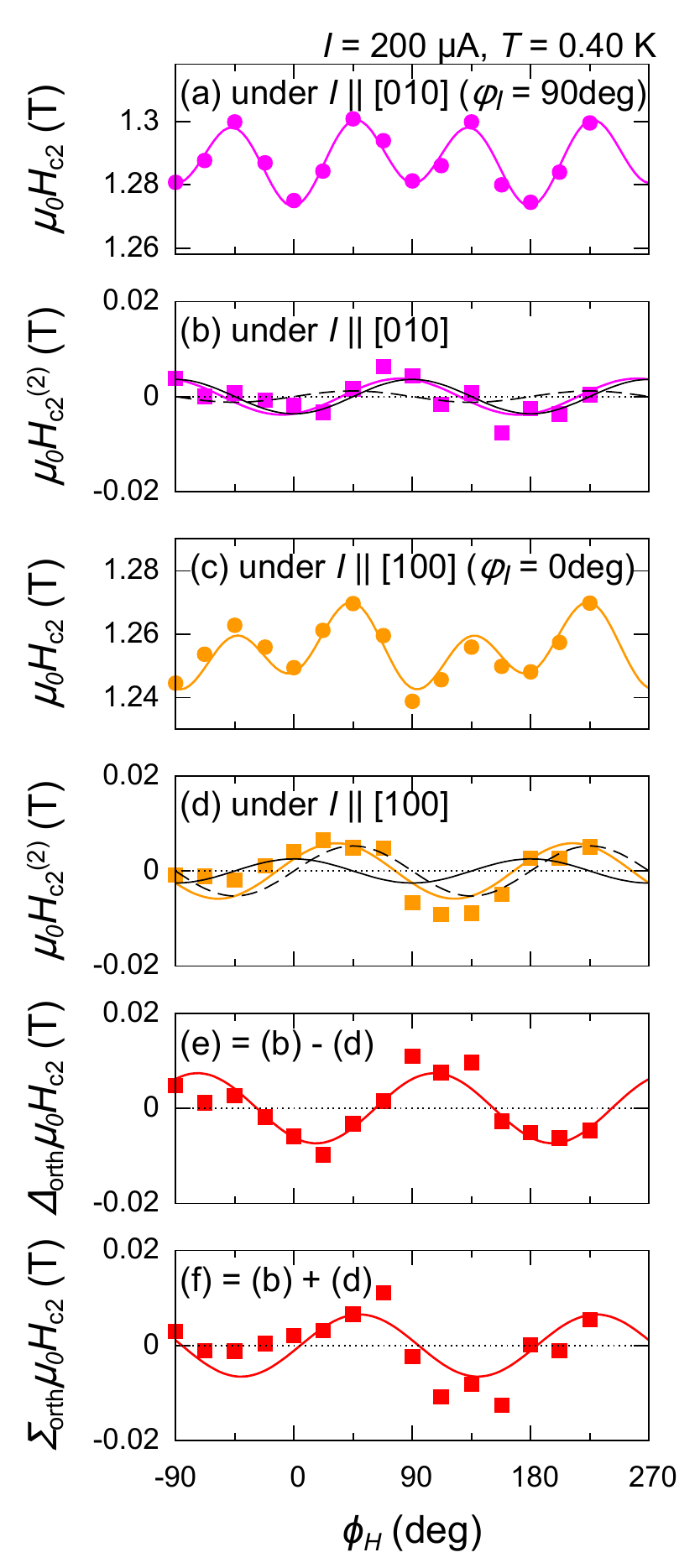}
\caption{$H_{\soeji{c2}}$ anisotropy of ${\rm Sr_2 Ru O_4}$ under in-plane current.
  Additional data are shown in SM \cite{SupplementalAraki}.
  (a) Raw $H_{\soeji{c2}}$ data as a function of $\phi_H$ under current along the [010] direction.
      The data are well fitted with a combination of four-fold and two-fold sinusoidal function (Eq.~(\ref{hc2})) as
      shown with the pink solid curve.
  (b) two-fold component of $H_{\soeji{c2}}$ under $I//[010]$ (pink squares),
   extracted from the raw $H_{\soeji{c2}}$ data by subtracting the fitted four-fold component and the constant offset.
  The pink curve present the two-fold sinusoidal fitting, whereas the 
   solid and broken black curves show the decomposed cosine and sine terms of the pink curve,
   respectively.
  (c) Raw $H_{\soeji{c2}}$ data (circles) and fitting result (solid curve) under current along the [100] direction.
  Compared to (a), the data suggest that current direction $\phi_I$ switches the sign of the difference between
   $H_{\soeji{c2}}$ at $\phi_H = 0^{\circ}$ and that at $\phi_H = 90^{\circ}$.
  (d) two-fold component of $H_{\soeji{c2}}$ under $I//[100]$ together with the two-fold fitting (orange curve)
  and the decomposed cosine (black solid curve) and sine (black broken curve) components.
  (e) $I$-induced two-fold component of $H_{\soeji{c2}}$ obtained by
 taking the difference between the data in (b) and (d).
 This only contains the two-fold component that is switched by the change of the current direction.
  (f) $I$-independent two-fold component of $H_{\soeji{c2}}$ obtained by
  taking the summation of the data in (b) and (d).
  This only contains the two-fold component that is not switched by the change of the current direction.
}\label{subtract}
\end{figure}

Figures~\ref{subtract}(a) and \ref{subtract}(c) compare the $\phi_H$ dependence of the in-plane 
 $H_{\soeji{c2}}$ for $I//[010]$ and $I//[100]$ measured at $200~\muup$A and $0.40$~K.
Under in-plane currents, $H_{\soeji{c2}}(\phi_H)$ clearly shows not only the known four-fold anisotropy
 but also an additional two-fold anisotropy.
This is also evident in the raw data in Fig.~\ref{082}:
For example, $H_{\soeji{c2}}$ for $\phi_H = 45^{\circ}$ is noticeably larger than that for $\phi_H = 135^{\circ}$,
  although these two conditions should be equivalent considering the tetragonal crystalline symmetry
  if the current were absent.
We have confirmed that this two-fold behavior is 
 not caused by the misalignment of fields with respect to the $ab$ plane (Fig.~S2 in SM \cite{SupplementalAraki}).
Moreover, comparing Figs.~\ref{subtract}(a) and (c), we notice that 
 this two-fold components of $H_{\soeji{c2}}$ are altered by changing the current direction $\phi_I$.
For example, the $H_{\soeji{c2}}$ value at $\phi_H = 0^{\circ}$ is smaller than that at $90^{\circ}$ for $I//[010]$ (Fig.~3(a)),
 whereas they are opposite for $I//[100]$ (Fig.~3(c)).
This fact also evidences that the external current plays an important role in the observed two-fold behavior.

To analyze the two-fold behavior quantitatively, 
 we fit the $H_{\soeji{c2}}(\phi_H)$ data with a combination of two and four-fold
 sinusoidal functions with a constant offset:
\begin{alignat}{1}
  \mu_0 h_{\soeji{c2}}(\phi_H) = a_0 + a_2 \cos 2\phi_H + b_2\sin 2\phi_H + a_4\cos 4\phi_H.  \label{hc2}
\end{alignat}
We comment that the four-fold cosine term represents the known $H_{\soeji{c2}}$ anisotropy under zero current
  ($H_{\soeji{c2//[110]}} > H_{\soeji{c2//[100]}}$) \cite{Kittaka2009.PhysRevB.80.174514,Mao2000.PhysRevLett.84.991}
  and we have checked that the four-fold sine term is negligible even under currents.
As exemplified by the solid curves in Figs.~\ref{subtract}(a) and (c), the fitting was
 successful for the $H_{\soeji{c2}}(\phi_H)$ data sets.
The obtained value of $a_4$ is $\sim -0.011$~T and $\sim -0.010$~T for the data in Fig.~3(a) and (c).
These values are consistent with the previous studies 
\cite{Mao2000.PhysRevLett.84.991,Kittaka2009.PhysRevB.80.174514}.
Notice that the quantity $\Delta(\mu_0 H_{\soeji{c2}})=\mu_0 H_{\soeji{c2}}[110] - \mu_0 H_{\soeji{c2}}[100]$
 in the literature corresponds to the quantity $-2a_4$ of our study.
Then, to extract the two-fold component $H_{\soeji{c2}}^{(2)}$,
 we subtracted the fitted four-fold component $a_4 \cos 4\phi_H$ and the offset $a_0$ from the data.
The results are shown in Figs.~\ref{subtract}(b) and (d).
Here, the colored curves represent the two-fold terms obtained by the fitting, whereas the black solid and broken curves are decomposed $\cos 2\phi_H$
 and $\sin 2\phi_H$ components, respectively.
Between $I//[100]$ and $[010]$, $H_{\soeji{c2}}^{(2)}$ is clearly different with the opposite sign of the cosine component.

\begin{figure} 
  \centering
  \includegraphics[keepaspectratio, scale=0.88]{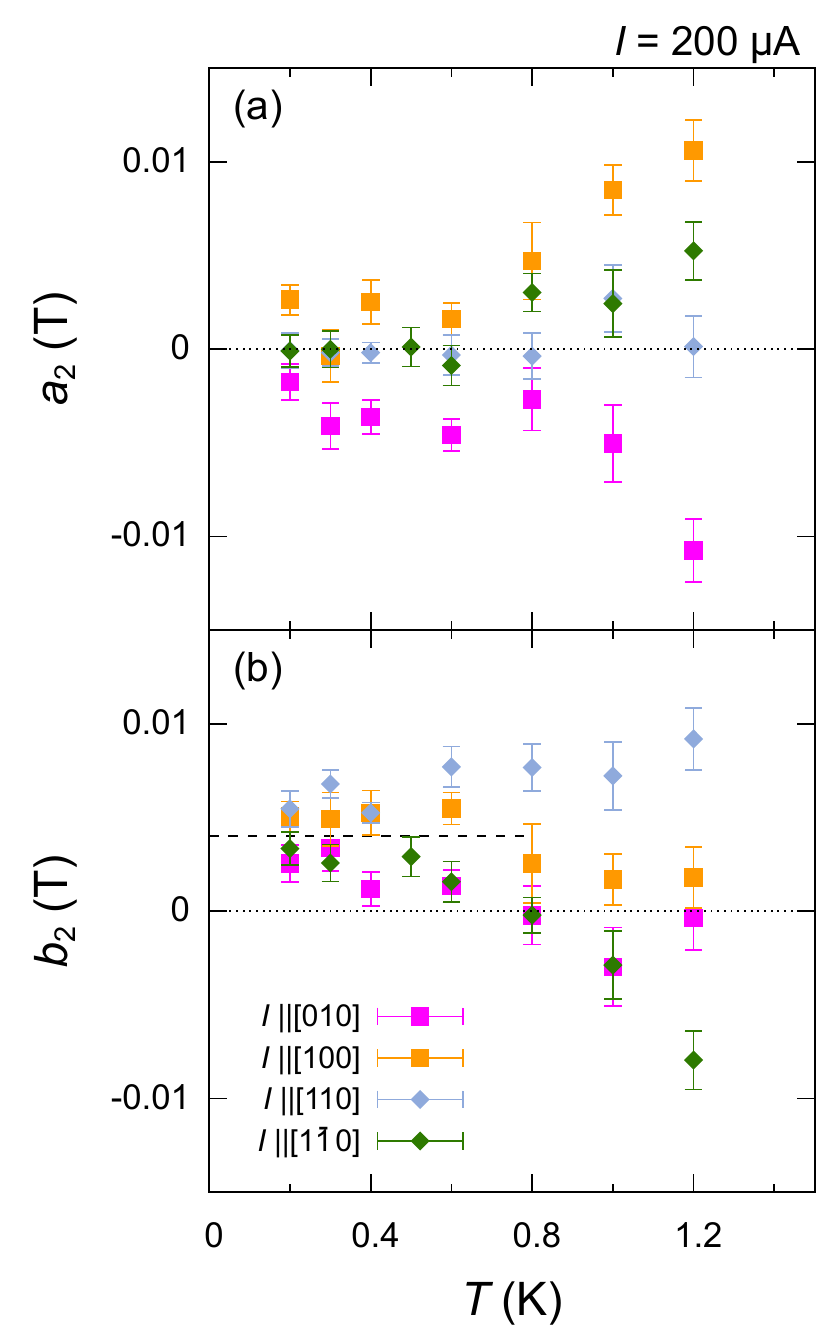}
\caption{
  Temperature dependence of $H_{\soeji{c2}}^{(2)}$ measured under current along the four directions.
  The pink, yellow, blue and green points correspond to [010],
  [100], [110] and $[1\overline{1}0]$ current directions.
  (a) Coefficient of the two-fold cosine component $a_2$. $a_2$ under $I//[100]$ and $[010]$ increases
   on warming and their signs are switched by current-directions.
   In contrast, $a_2$ under $I//[110]$ and $[1\overline{1}0]$ stays nearly zero.
  These results suggest that $H_{\soeji{c2}}^{(2)}$ is mostly induced by current and thus behaves as
   $\cos 2(\phi_H - \phi_I)$.
  (b) Coefficient of the two-fold sine component $b_2$.
   Similar to $a_2$, the current-direction dependence of $b_2$ at high temperature is consistent with
   the expectation $H_{\soeji{c2}}^{(2)} \propto \cos 2(\phi_H -\phi_I)$.
   However, unlike $a_2$, $b_2$ under all current-directions have an additional term
   reaching nearly 0.005 T on cooling to 0~K.
}\label{a2b2}
\end{figure}

We present the $T$-dependence of the obtained fitting parameters $a_2$ and $b_2$ in eq.~{\ref{hc2}}
 for various current directions in Fig.~\ref{a2b2}. (Behavior of $a_0$ and $a_4$ will be discussed later.)
For the coefficient of the $\cos 2\phi_H$ component $a_2$, its magnitude under the orthogonal current directions, namely
 $I//[100]$ and $I//[010]$, increases on warming while its sign is switched by current-direction change.
This is indeed expected for a current-induced effect, where by symmetry $H_{\soeji{c2}}^{(2)}$ should
 behave as $\propto \cos 2(\phi_H -\phi_I) = \cos 2\phi_H$ for $\phi_I = 0^{\circ} ~(I//[100])$
 and $-\cos 2\phi_H$ for $\phi_I = 90^{\circ} ~(I//[010])$.
In contrast, $a_2$ under the diagonal current directions, namely $I//[110]$ and $I//[1\overline{1}0]$,
 are nearly independent of temperature and stays close to zero.
This is again consistent with the $\cos 2(\phi_H -\phi_I)$ behavior for the current-induced $H_{\soeji{c2}}^{(2)}$,
 since $\cos 2(\phi_H -\phi_I) = \pm \sin 2\phi_H$ for $\phi_I = \pm 45^{\circ}$ and thus the cosine component
 should be zero.
Figure~\ref{a2b2}(b) shows the temperature dependence of the coefficients $b_2$. 
At high temperatures, the behavior of $b_2$ is similar to that of $a_2$ 
 except for an exchange of the roles between the orthogonal and diagonal current directions:
  $b_2$ is finite and its sign depends on the current direction $I$//[110] or $[1\overline{1}0]$,
   but is nearly zero for $I$//[100] and [010].
Thus this high-temperature behavior in $b_2$ again indicate $H_{\soeji{c2}}^{(2)} \propto \cos 2(\phi_H -\phi_I)$ behavior.
However, at lower temperatures, $b_2$ tends to converge to a finite value $b_2(T \to 0)\sim 0.005$~T
 irrespective of the current directions.
Such convergence reveals an additional two-fold component that
 is not caused by external current.


From these analyses, we revealed that the $H_{c2}^{(2)}$ data have current-independent and current-dependent contributions,
 which we hereafter express as $H_{\soeji{c2}}^{\soeji{(2ind)}}$ and $H_{\soeji{c2}}^{\soeji{(2dep)}}$.
The former corresponds to the low-temperature current-direction independent behavior in $b_{2}$ (Fig.~\ref{a2b2}(b)) and
 the latter corresponds to the $\cos 2(\phi_H -\phi_I)$ term as seen in the behavior common to $a_2$ and $b_2$
 discussed in the previous paragraph.  
Further analyses and interpretations of these contributions will be provided in the next section.


Before closing this section, we comment on possible heating effects due to current.
We estimate the upper limit of the actual sample temperature from the current dependence of the coefficient $a_0$
 by assuming that its variation with current is solely due to Joule heating (see SM \cite{SupplementalAraki}).
This estimation indicates that the temperature increase is at most 0.2~K in the lowest temperature region and less than 0.1~K
 above 0.4~K.


\section{Discussion}

As explained in the previous section, we extracted two-fold anisotropy of $H_{\soeji{c2}}$ of ${\rm Sr_2 Ru O_4}$
 under current.
We reveal systematically how this additional anisotropy depends on current-strength, current-directions, 
 and temperature.
As a well-known effect, external current induces vortex flows in the mixed state by the Lorentz force,
 leading to the apparent reduction in $H_{\soeji{c2}}$.
Thus, such an effect is maximized under field perpendicular to current
 and would not occur under field parallel to current;
 As a result, this effect is expressed as a two-fold cosine-like component $\propto \cos2(\phi_H - \phi_I)$,
 whose sign is negative under field perpendicular to current (i.e. $\phi_H - \phi_I = \pm90^{\circ}$)
 and positive under field parallel to current (i.e. $\phi_H = \phi_I$).

For ordinary type-II superconductors (both 2D and 3D) in which the main pair-breaking effect is the ordinary
 orbital effect, the vortex-flow effect is expected to be observed.
However, in ${\rm Sr_2 Ru O_4}$, the FOT occurs below 0.8~K,
 indicating that the dominated pair-breaking effect is not the orbital effect.
Thus, the two-fold cosine-like component can behave differently between the FOT
 and the SOT regions.


In order to extract the current-dependent term including the vortex-flow effect from the $H_{\soeji{c2}}^{(2)}$ data,
 we take the difference between the $H_{\soeji{c2}}^{(2)}(\phi_H, \phi_I=90^{\circ})$ data and 
 $H_{\soeji{c2}}^{(2)}(\phi_H, \phi_I=0^{\circ})$ data.
This process is motivated by the following expectation.
We expect that the two-fold component $H_{\soeji{c2}}^{(2)}(\phi_H, \phi_I)$
contain current-independent term $H_{\soeji{c2}}^{(2\soeji{ind})}(\phi_H)$ 
and current-dependent term $H_{\soeji{c2}}^{(2\soeji{dep})}(\phi_H -\phi_I)$, $i.e.$ 
\begin{alignat}{1}
  H_{\soeji{c2}}^{(2)}(\phi_H, \phi_I) = H_{\soeji{c2}}^{(2\soeji{ind})}(\phi_H) + H_{\soeji{c2}}^{(2\soeji{dep})}(\phi_H -\phi_I). 
\end{alignat}
Then, due to the two-fold nature, $H_{\soeji{c2}}^{(2\soeji{dep})}$ should change sign upon a 
 current-direction switching of 90$^{\circ}$, as
\begin{alignat}{1}
  H_{\soeji{c2}}^{(2\soeji{dep})}(\phi_H -(\phi_I+90^{\circ})) = -H_{\soeji{c2}}^{(2\soeji{dep})}(\phi_H -\phi_I).
\end{alignat}  
In contrast, $H_{\soeji{c2}}^{(2\soeji{ind})}$ is, by definition, independent of the 90$^{\circ}$ current direction
 switching.
Thus, by calculating 
\begin{alignat}{1}
  \varDelta_{\soeji{orth}}H_{\soeji{c2}}^{\soeji{(2)}} & \equiv H_{\soeji{c2}}^{\soeji{(2)}}
(\phi_I = 90^{\circ}) - H_{\soeji{c2}}^{\soeji{(2)}}(\phi_I = 0^{\circ}), 
\end{alignat}
 $H_{\soeji{c2}}^{\soeji{(2ind)}}$ is eliminated and we can expect that 
 $\varDelta_{\soeji{orth}}H_{\soeji{c2}}^{(2)} \simeq 2H_{\soeji{c2}}^{(2\soeji{dep})}(\phi_I = 90^{\circ})$.
A representative $\varDelta_{\soeji{orth}}H_{\soeji{c2}}^{\soeji{(2)}}$ is shown in Fig.~\ref{subtract}(e).
Similarly, 
\begin{alignat}{1}
  \varDelta_{\soeji{diag}}H_{\soeji{c2}}^{\soeji{(2)}} & \equiv H_{\soeji{c2}}^{\soeji{(2)}}
(\phi_I = 45^{\circ}) - H_{\soeji{c2}}^{\soeji{(2)}}(\phi_I = -45^{\circ}) 
\end{alignat}
should satisfy $\varDelta_{\soeji{diag}}H_{\soeji{c2}}^{(2)} \simeq 2H_{\soeji{c2}}^{(2\soeji{dep})}$ (Fig.~S4), 
 providing an independent evaluation of $H_{\soeji{c2}}^{(2\soeji{dep})}(\phi_I = 45^{\circ})$.
On the other hand, we can extract the current-independent term $H_{\soeji{c2}}^{(2\soeji{ind})}(\phi_H)$
 by evaluating sums of the datasets with 90$^{\circ}$ current switching: 
\begin{alignat}{1}
  \varSigma_{\soeji{orth}}H_{\soeji{c2}}^{\soeji{(2)}} & \equiv H_{\soeji{c2}}^{\soeji{(2)}}
(\phi_I = 90^{\circ}) + H_{\soeji{c2}}^{\soeji{(2)}}(\phi_I = 0^{\circ})
\end{alignat}
and
\begin{alignat}{1}
  \varSigma_{\soeji{diag}}H_{\soeji{c2}}^{\soeji{(2)}} & \equiv H_{\soeji{c2}}^{\soeji{(2)}}
(\phi_I = 45^{\circ}) + H_{\soeji{c2}}^{\soeji{(2)}}(\phi_I = -45^{\circ}).
\end{alignat}
In these sums, current-dependent terms are eliminated and we expect $\varSigma_{\soeji{orth}}H_{\soeji{c2}}^{(2)}
 \simeq \varSigma_{\soeji{diag}}H_{\soeji{c2}}^{(2)} \simeq 2H_{\soeji{c2}}^{(2\soeji{ind})}$.
A representative $\varSigma_{\soeji{orth}}H_{\soeji{c2}}^{\soeji{(2)}}$
 is shown in Fig.~\ref{subtract}(f) and $\varSigma_{\soeji{diag}}H_{\soeji{c2}}^{\soeji{(2)}}$
 in Fig.~S4.

\begin{figure}
  \centering
  \includegraphics[keepaspectratio, scale=0.90]{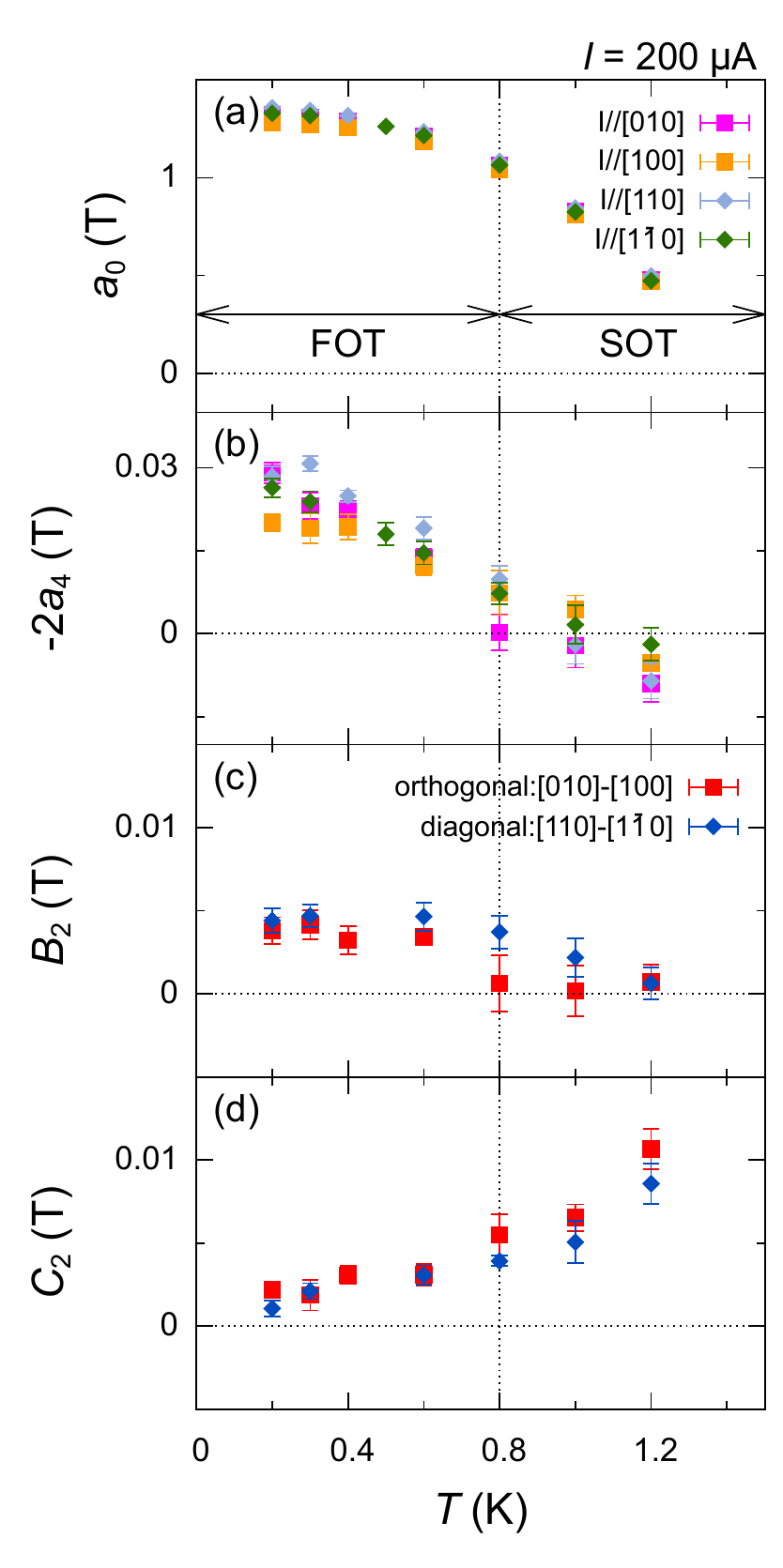}
     \caption{
      Relation among the first-order transition (FOT) region and $H_{\soeji{c2}}$ anisotropies under current.
      (a) $T$-dependence of $a_0$.
      Since $a_0$ corresponds to the in-plane average value of $H_{\soeji{c2}}$,
      this curve provide the $H_{\soeji{c2}}$-$T$ phase diagram of this sample.
      (b) $T$-dependence of the four-fold component $-2a_4$.
      The behavior of $-2a_4$ is consistent with the previous studies~\cite{Mao2000.PhysRevLett.84.991,Kittaka2009.PhysRevB.80.174514}.
      In (a) and (b), the pink, orange, sky-blue and green points are for the data obtained 
       with $I$//[010], [100], [110] and $[1\overline{1}0]$, respectively.
      (c) and (d); $T$-dependence of two-fold $I$-independent component and $I$-dependent component
       of $H_{\soeji{c2}}^{(2)}$.
      Red squares and blue diamonds are the data obtained for orthogonal and diagonal current-directions, respectively.
      The two-fold $I$-independent sine component $B_2$ reaches about 0.005 T
       in the FOT region ($T\leq 0.8$~K) and nearly zero in the second-order transition region ($T\geq 0.8$~K).
      This result indicates that the sample in a high-field state becomes more sensitive 
       to inhomogeneity effect, producing two-fold anisotropy.
      The $\cos 2(\phi_H -\phi_I)$ component $C_2$ in (d) increase on warming. This observation
       is consistent with vortex flow effect.
     }
     \label{Capital}
\end{figure}

In Figs.~\ref{subtract}(e), \ref{subtract}(f), S4, one can clearly see that $\varDelta_{\soeji{orth}}H_{\soeji{c2}}^{\soeji{(2)}}$
 and $\varDelta_{\soeji{diag}}H_{\soeji{c2}}^{\soeji{(2)}}$ are dominated by $\cos2(\phi_H-\phi_I)
 |_{\phi_I = 90^{\circ}, 45^{\circ}}$
 and $\varSigma_{\soeji{orth}}H_{\soeji{c2}}^{\soeji{(2)}}$
 and $\varSigma_{\soeji{diag}}H_{\soeji{c2}}^{\soeji{(2)}}$ are by $\sin2\phi_H$.
To perform quantitative analysis, we fitted $(1/2)\varSigma H_{\soeji{c2}}^{(2)}$ by $A_2\cos2\phi_H + B_2\sin2\phi_H$
 and $(1/2)\varDelta H_{\soeji{c2}}^{(2)}$ by $C_2\cos2(\phi_H-\phi_I)|_{\phi_I = 90^{\circ}, 45^{\circ}}
 + D_2\sin2(\phi_H-\phi_I)|_{\phi_I = 90^{\circ}, 45^{\circ}}$.
The resultant parameters are plotted in Figs.~\ref{Capital}(c), (d), S6(a) and S6(b).
We confirmed that the coefficients evaluated from the orthogonal and diagonal combinations
 agree with each other, manifesting the validity of the analyses.
We then indeed find $A_2$ and $D_2$ are small and do not exhibit temperature dependence.
In contrast, $B_2$ and $C_2$ plotted in Figs.~\ref{Capital}(c) and (d) are much larger than $A_2$ or $D_2$
 and exhibits characteristic temperature dependence as will be discussed later.  



In Figs.~\ref{Capital}(a) and (b) the angle-independent component $a_0$ (see Eq.~(\ref{hc2}))
 and four-fold component $-2a_4$ are compared with the coefficients $B_2$ and $C_2$.
Notice that  the coefficient $a_0$ in Eq.~(\ref{hc2}) can be considered as the in-plane average $H_{\soeji{c2}}$
 and thus provide the $H_{\soeji{c2}}$-$T$ phase diagram of ${\rm Sr_2RuO_4}$.
Indeed, Fig.~\ref{Capital}(a) agrees well with the phase diagram reported in literature~\cite{Kittaka2009.PhysRevB.80.174514,Yonezawa2014.JPhysSocJpn.83.083706}.
The value $-2a_4$ should be equal to $\mu_0H_{\soeji{c2//[110]}}-\mu_0H_{\soeji{c2//[100]}}$
 in the absence of current, characterizing the known four-fold $H_{\soeji{c2}}$ anisotropy.
This quantity is nearly zero above 0.8~K but increase on cooling in the FOT region and reaches 0.03~T at the lowest temperature.
Such behavior is consistent with the previous studies.

As shown in Fig.~\ref{Capital}(d),
the coefficient $C_2$ exhibits a finite value and decreases on cooling.
The coefficient $D_2$ exhibits a smaller value than $C_2$ and
 show only a weak temperature dependence (see SM \cite{SupplementalAraki}).
These results indicate that the current-dependent components are cosine-like and increase on warming.
It is consistent with the scenario where vortex flow gives current-dependent two-fold anisotropy of $H_{\soeji{c2}}$.

The coefficient $B_2$ shown in Fig.~\ref{Capital}(c), characterizing current-independent two-fold behavior,
 is finite only below 0.8~K in the FOT region, for the orthogonal current.
A similar tendency is clear for the diagonal current.
Since other bulk measurements have not revealed such a current-independent two-fold component
 of $H_{\soeji{c2}}$,
 the behavior of $B_{2}$ is probably related to minute surface damage caused by the FIB processing.
However, it is not sufficient to explain the $T$-dependence of $B_2$
 because it does not change smoothly but appear only in the FOT region.
This behavior can be explained if the resistivity suddenly becomes
 more sensitive to the inhomogeneity in the FOT region.
One possible candidate for such sudden change in the resistivity property is
 the formation of the FFLO state, which is accompanied by a real-space modulation
 of the superconducting order parameter and is recently revealed by an NMR experiment~\cite{Kinjoimplement}.
 If there is a surface damage, for example, the FFLO order-parameter modulation can be pinned to the surface
 damage and such pinning can induce additional two-fold behavior in $H_{\soeji{c2}}(\phi_H)$.

\section{Summary and conclusion}
Here we summarize the main results of our study.

  (1) The angle-independent component $a_0$ in Eq.~(\ref{hc2}) can be considered as the in-plane average $H_{\soeji{c2}}$.

  (2) The quantity $-2a_4$ characterizes the known four-fold $H_{\soeji{c2}}$ anisotropy.
It corresponds to $\varDelta(\mu_0H_{\soeji{c2}})$ in the previous studies \cite{Mao2000.PhysRevLett.84.991,Kittaka2009.PhysRevB.80.174514}
 and exhibits about 80$\%$ of the value of $\varDelta(\mu_0H_{\soeji{c2}})$ in the first-order transition region. 
In the second-order transition region, $-2a_4$ shows close to zero in the same way as $\varDelta(\mu_0H_{\soeji{c2}})$.
Thus, the four-fold anisotropy in $H_{\soeji{c2}}$ under in-plane current shows the same $T$-dependence
 as that of the four-fold anisotropy in $H_{\soeji{c2}}$ under zero current.

  (3) The coefficient $C_2$ shows the $\cos 2(\phi_H -\phi_I)$ component of $H_{\soeji{c2}}^{(2\soeji{dep})}$.
Since it takes positively finite values and increases with temperature, 
 the current-induced two-fold anisotropy in $H_{\soeji{c2}}^{(2)}$ is probably due to the vortex-flow effect.

  (4) The coefficient $B_2$ shows the two-fold $I$-independent sine component of $H_{\soeji{c2}}^{(2)}$.  
Such a two-fold component has not been observed in previous experiments \cite{Yonezawa2014.JPhysSocJpn.83.083706,Mao2000.PhysRevLett.84.991,Kittaka2009.PhysRevB.80.174514}.
This quantity revealed in our study becomes finite only in the first-order transition region.
This is probably related to minute the crystal distortion and surface damage caused by FIB processing.
However, it is not enough to explain why $B_2$ becomes finite only in the first-order transition region.
As a possible scenario, we propose that the increased sensitivity to inhomogeneity is due to the realization of FFLO state
 in the first-order transition region.

In conclusion, we have established a clear approach to probe the electrical resistivity
 of two-dimensional superconductors, such as ${\rm Sr_2 Ru O_4}$, with a varying current and magnetic field directions.
The combination of the vector magnet and our FIB structured sample
 enables us to control the azimuthal field angle $\phi_{H}$,
 and switch in-plane current-direction $\phi_I$ along four crystal orientations.
We have demonstrated that $H_{\soeji{c2}}$ in the $ab$ plane of ${\rm Sr_2 Ru O_4}$ exhibits 
 not only an ordinary four-fold anisotropy but also an additional two-fold anisotropy.
This two-fold anisotropy increases on warming and probably expressed as $\cos 2(\phi_H - \phi_I)$,
 suggesting that flux flow resistivity generated by the Lorentz effect
 leads to underestimation of $H_{\soeji{c2}}$ in the $ab$ plane.
We find another two-fold term behaving as $\sin2\phi_H$, independent of the current direction.
This component, indicative of the sensitivity of the superconductivity to sample inhomogenity,
 becomes noticible only below 0.8~K, where field-induced phases such as the FFLO state are highly anticipated.

Our study is expected to provide hints toward clarification of the remaining issues
 on ${\rm Sr_2 Ru O_4}$, such as the origin of the enhancement of $\Tc$ under uniaxial strain
 or in eutectic crystals,
 and the relationship between the superconducting order parameters and the
 anisotropy of $H_{\soeji{c2}}$. 
In addition, our resistivity measurement system will be helpful for investigating
 interesting phenomena in other materials such as coupling between current and electronic nematicity in 
 nematic superconductors \cite{Yonezawa2019condmat4010002} or in nematic electron liquid systems \cite{Fernandes2014nematicorder}.


%

\section{acknowledgement}

  The authors thank Y. Yanase, K. Machida, K. Ishida, S. Kitagawa, G. Mattoni, for valuable discussions.
  We also acknowledge technical supports from Kyoto Univ. LTM Center.
  This work is supported by
  Japan Society for the Promotion of Science (JSPS) Core-to-core program (No. JPJSCCA20170002)
  and by JSPS KAKENHI Nos. JP15H05852, JP15K21717, JP17H06136, and JP20H05158.
  This work is also supported by Research Grants Council of Hong Kong (CUHK 14301316) and CUHK Direct Grants (No. 4053299, 4053345).


\begin{thebibliography}{32}%
  \makeatletter
  \providecommand \@ifxundefined [1]{%
   \@ifx{#1\undefined}
  }%
  \providecommand \@ifnum [1]{%
   \ifnum #1\expandafter \@firstoftwo
   \else \expandafter \@secondoftwo
   \fi
  }%
  \providecommand \@ifx [1]{%
   \ifx #1\expandafter \@firstoftwo
   \else \expandafter \@secondoftwo
   \fi
  }%
  \providecommand \natexlab [1]{#1}%
  \providecommand \enquote  [1]{``#1''}%
  \providecommand \bibnamefont  [1]{#1}%
  \providecommand \bibfnamefont [1]{#1}%
  \providecommand \citenamefont [1]{#1}%
  \providecommand \href@noop [0]{\@secondoftwo}%
  \providecommand \href [0]{\begingroup \@sanitize@url \@href}%
  \providecommand \@href[1]{\@@startlink{#1}\@@href}%
  \providecommand \@@href[1]{\endgroup#1\@@endlink}%
  \providecommand \@sanitize@url [0]{\catcode `\\12\catcode `\$12\catcode
    `\&12\catcode `\#12\catcode `\^12\catcode `\_12\catcode `\%12\relax}%
  \providecommand \@@startlink[1]{}%
  \providecommand \@@endlink[0]{}%
  \providecommand \url  [0]{\begingroup\@sanitize@url \@url }%
  \providecommand \@url [1]{\endgroup\@href {#1}{\urlprefix }}%
  \providecommand \urlprefix  [0]{URL }%
  \providecommand \Eprint [0]{\href }%
  \@ifxundefined \urlstyle {%
    \providecommand \doi  [0]{\begingroup \@sanitize@url \@doi}%
    \providecommand \@doi [1]{\endgroup \@@startlink {\doibase
    #1}doi:\discretionary {}{}{}#1\@@endlink }%
  }{%
    \providecommand \doi  [0]{doi:\discretionary{}{}{}\begingroup
    \urlstyle{rm}\Url }%
  }%
  \providecommand \doibase [0]{http://dx.doi.org/}%
  \providecommand \Doi [0]{\begingroup \@sanitize@url \@Doi }%
  \providecommand \@Doi  [1]{\endgroup\@@startlink{\doibase#1}\@@Doi}%
  \providecommand \@@Doi [1]{#1\@@endlink}%
  \providecommand \selectlanguage [0]{\@gobble}%
  \providecommand \bibinfo  [0]{\@secondoftwo}%
  \providecommand \bibfield  [0]{\@secondoftwo}%
  \providecommand \translation [1]{[#1]}%
  \providecommand \BibitemOpen [0]{}%
  \providecommand \bibitemStop [0]{}%
  \providecommand \bibitemNoStop [0]{.\EOS\space}%
  \providecommand \EOS [0]{\spacefactor3000\relax}%
  \providecommand \BibitemShut  [1]{\csname bibitem#1\endcsname}%
  \bibitem [{\citenamefont {Maeno}\ \emph {et~al.}(1994)\citenamefont {Maeno},
    \citenamefont {Hashimoto}, \citenamefont {Yoshida}, \citenamefont
    {Nishizaki}, \citenamefont {Fujita}, \citenamefont {Bednorz},\ and\
    \citenamefont {Lichtenberg}}]{Maeno1994}%
    \BibitemOpen
    \bibfield  {author} {\bibinfo {author} {\bibfnamefont {Y.}~\bibnamefont
    {Maeno}}, \bibinfo {author} {\bibfnamefont {H.}~\bibnamefont {Hashimoto}},
    \bibinfo {author} {\bibfnamefont {K.}~\bibnamefont {Yoshida}}, \bibinfo
    {author} {\bibfnamefont {S.}~\bibnamefont {Nishizaki}}, \bibinfo {author}
    {\bibfnamefont {T.}~\bibnamefont {Fujita}}, \bibinfo {author} {\bibfnamefont
    {J.~G.}\ \bibnamefont {Bednorz}},\ and\ \bibinfo {author} {\bibfnamefont
    {F.}~\bibnamefont {Lichtenberg}},\ }\bibfield  {title} {\enquote {\bibinfo
    {title} {Superconductivity in a layered perovskite without copper}}\
    }\href@noop {} {\bibfield  {journal} {\bibinfo  {journal} {Nature}\ }\textbf
    {\bibinfo {volume} {372}},\ \bibinfo {pages} {532} (\bibinfo {year}
    {1994})}\BibitemShut {NoStop}%
  \bibitem [{\citenamefont {Bergemann}\ \emph {et~al.}(2003)\citenamefont
    {Bergemann}, \citenamefont {Mackenzie}, \citenamefont {Julian}, \citenamefont
    {Forsythe},\ and\ \citenamefont {Ohmichi}}]{Bergemann2003.AdvPhys.52.639}%
    \BibitemOpen
    \bibfield  {author} {\bibinfo {author} {\bibfnamefont {C.}~\bibnamefont
    {Bergemann}}, \bibinfo {author} {\bibfnamefont {A.~P.}\ \bibnamefont
    {Mackenzie}}, \bibinfo {author} {\bibfnamefont {S.~R.}\ \bibnamefont
    {Julian}}, \bibinfo {author} {\bibfnamefont {D.}~\bibnamefont {Forsythe}},\
    and\ \bibinfo {author} {\bibfnamefont {E.}~\bibnamefont {Ohmichi}},\
    }\bibfield  {title} {\enquote {\bibinfo {title} {Quasi-two-dimensional Fermi
    liquid properties of the unconventional superconductor Sr$_2$RuO$_4$}}\ }\Doi
    {0.1080/00018730310001621737} {\bibfield  {journal} {\bibinfo  {journal}
    {Adv. Phys.}\ }\textbf {\bibinfo {volume} {52}},\ \bibinfo {pages} {639}
    (\bibinfo {year} {2003})}\BibitemShut {NoStop}%
  \bibitem [{\citenamefont {Damascelli}\ \emph {et~al.}(2000)\citenamefont
    {Damascelli}, \citenamefont {Lu}, \citenamefont {Shen}, \citenamefont
    {Armitage}, \citenamefont {Ronning}, \citenamefont {Feng}, \citenamefont
    {Kim}, \citenamefont {Shen}, \citenamefont {Kimura}, \citenamefont {Tokura},
    \citenamefont {Mao},\ and\ \citenamefont
    {Maeno}}]{Damascelli2000.PhysRevLett.85.5194}%
    \BibitemOpen
    \bibfield  {author} {\bibinfo {author} {\bibfnamefont {A.}~\bibnamefont
    {Damascelli}}, \bibinfo {author} {\bibfnamefont {D.~H.}\ \bibnamefont {Lu}},
    \bibinfo {author} {\bibfnamefont {K.~M.}\ \bibnamefont {Shen}}, \bibinfo
    {author} {\bibfnamefont {N.~P.}\ \bibnamefont {Armitage}}, \bibinfo {author}
    {\bibfnamefont {F.}~\bibnamefont {Ronning}}, \bibinfo {author} {\bibfnamefont
    {D.~L.}\ \bibnamefont {Feng}}, \bibinfo {author} {\bibfnamefont
    {C.}~\bibnamefont {Kim}}, \bibinfo {author} {\bibfnamefont {Z.-X.}\
    \bibnamefont {Shen}}, \bibinfo {author} {\bibfnamefont {T.}~\bibnamefont
    {Kimura}}, \bibinfo {author} {\bibfnamefont {Y.}~\bibnamefont {Tokura}},
    \bibinfo {author} {\bibfnamefont {Z.~Q.}\ \bibnamefont {Mao}},\ and\ \bibinfo
    {author} {\bibfnamefont {Y.}~\bibnamefont {Maeno}},\ }\bibfield  {title}
    {\enquote {\bibinfo {title} {Fermi Surface, Surface States, and Surface
    Reconstruction in Sr$_2$RuO$_4$}}\ }\Doi {10.1103/PhysRevLett.85.5194}
    {\bibfield  {journal} {\bibinfo  {journal} {Phys. Rev. Lett.}\ }\textbf
    {\bibinfo {volume} {85}},\ \bibinfo {pages} {5194} (\bibinfo {year}
    {2000})}\BibitemShut {NoStop}%
  \bibitem [{\citenamefont {Pustogow}\ \emph {et~al.}(2019)\citenamefont
    {Pustogow}, \citenamefont {Luo}, \citenamefont {Chronister}, \citenamefont
    {Su}, \citenamefont {Sokolov}, \citenamefont {Jerzembeck}, \citenamefont
    {Mackenzie}, \citenamefont {Hicks}, \citenamefont {Kikugawa}, \citenamefont
    {Raghu}, \citenamefont {Bauer},\ and\ \citenamefont
    {Brown}}]{Pustogow2019.Nature.574.72}%
    \BibitemOpen
    \bibfield  {author} {\bibinfo {author} {\bibfnamefont {A.}~\bibnamefont
    {Pustogow}}, \bibinfo {author} {\bibfnamefont {Y.}~\bibnamefont {Luo}},
    \bibinfo {author} {\bibfnamefont {A.}~\bibnamefont {Chronister}}, \bibinfo
    {author} {\bibfnamefont {Y.-S.}\ \bibnamefont {Su}}, \bibinfo {author}
    {\bibfnamefont {D.~A.}\ \bibnamefont {Sokolov}}, \bibinfo {author}
    {\bibfnamefont {F.}~\bibnamefont {Jerzembeck}}, \bibinfo {author}
    {\bibfnamefont {A.~P.}\ \bibnamefont {Mackenzie}}, \bibinfo {author}
    {\bibfnamefont {C.~W.}\ \bibnamefont {Hicks}}, \bibinfo {author}
    {\bibfnamefont {N.}~\bibnamefont {Kikugawa}}, \bibinfo {author}
    {\bibfnamefont {S.}~\bibnamefont {Raghu}}, \bibinfo {author} {\bibfnamefont
    {E.~D.}\ \bibnamefont {Bauer}},\ and\ \bibinfo {author} {\bibfnamefont
    {S.~E.}\ \bibnamefont {Brown}},\ }\bibfield  {title} {\enquote {\bibinfo
    {title} {Constraints on the superconducting order parameter in Sr$_2$RuO$_4$
    from oxygen-17 nuclear magnetic resonance}}\ }\Doi
    {10.1038/s41586-019-1596-2} {\bibfield  {journal} {\bibinfo  {journal}
    {Nature}\ }\textbf {\bibinfo {volume} {574}},\ \bibinfo {pages} {72}
    (\bibinfo {year} {2019})}\BibitemShut {NoStop}%
  \bibitem [{\citenamefont {Ishida}\ \emph {et~al.}(2020)\citenamefont {Ishida},
    \citenamefont {Manago}, \citenamefont {Kinjo},\ and\ \citenamefont
    {Maeno}}]{Ishida2020.JPSJ.89.034712}%
    \BibitemOpen
    \bibfield  {author} {\bibinfo {author} {\bibfnamefont {K.}~\bibnamefont
    {Ishida}}, \bibinfo {author} {\bibfnamefont {M.}~\bibnamefont {Manago}},
    \bibinfo {author} {\bibfnamefont {K.}~\bibnamefont {Kinjo}},\ and\ \bibinfo
    {author} {\bibfnamefont {Y.}~\bibnamefont {Maeno}},\ }\bibfield  {title}
    {\enquote {\bibinfo {title} {Reduction of the $^{17}$O Knight Shift in the
    Superconducting State and the Heat-up Effect by NMR Pulses on
    Sr$_2$RuO$_4$}}\ }\Doi {10.7566/JPSJ.89.034712} {\bibfield  {journal}
    {\bibinfo  {journal} {Journal of the Physical Society of Japan}\ }\textbf
    {\bibinfo {volume} {89}},\ \bibinfo {pages} {034712} (\bibinfo {year}
    {2020})}\BibitemShut {NoStop}%
  \bibitem [{\citenamefont {Chronister}\ \emph {et~al.}(2021)\citenamefont
    {Chronister}, \citenamefont {Pustogow}, \citenamefont {Kikugawa},
    \citenamefont {Sokolov}, \citenamefont {Jerzembeck}, \citenamefont
    {W.~Hicks}, \citenamefont {P.~Mackenzie}, \citenamefont {D.~Bauer},\ and\
    \citenamefont {E.~Brown}}]{Chronister2021.PNAS.118.25}%
    \BibitemOpen
    \bibfield  {author} {\bibinfo {author} {\bibfnamefont {A.}~\bibnamefont
    {Chronister}}, \bibinfo {author} {\bibfnamefont {A.}~\bibnamefont
    {Pustogow}}, \bibinfo {author} {\bibfnamefont {N.}~\bibnamefont {Kikugawa}},
    \bibinfo {author} {\bibfnamefont {D.~A.}\ \bibnamefont {Sokolov}}, \bibinfo
    {author} {\bibfnamefont {F.}~\bibnamefont {Jerzembeck}}, \bibinfo {author}
    {\bibfnamefont {C.}~\bibnamefont {W.~Hicks}}, \bibinfo {author}
    {\bibfnamefont {A.}~\bibnamefont {P.~Mackenzie}}, \bibinfo {author}
    {\bibfnamefont {E.}~\bibnamefont {D.~Bauer}},\ and\ \bibinfo {author}
    {\bibfnamefont {S.}~\bibnamefont {E.~Brown}},\ }\bibfield  {title} {\enquote
    {\bibinfo {title} {Evidence for even parity unconventional superconductivity
    in Sr$_2$RuO$_4$}}\ }\Doi {10.1073/pnas.2025313118} {\bibfield  {journal}
    {\bibinfo  {journal} {Proceedings of the National Academy of Sciences of the
    United States of America}\ }\textbf {\bibinfo {volume} {118}},\ \bibinfo
    {pages} {25} (\bibinfo {year} {2021})}\BibitemShut {NoStop}%
  \bibitem [{\citenamefont {Petsch}\ \emph {et~al.}(2020)\citenamefont {Petsch},
    \citenamefont {Zhu}, \citenamefont {Enderle}, \citenamefont {Mao},
    \citenamefont {Maeno}, \citenamefont {Mazin},\ and\ \citenamefont
    {Hayden}}]{Petsch2020.Phys.Rev.Lett.125.217004}%
    \BibitemOpen
    \bibfield  {author} {\bibinfo {author} {\bibfnamefont {A.~N.}\ \bibnamefont
    {Petsch}}, \bibinfo {author} {\bibfnamefont {M.}~\bibnamefont {Zhu}},
    \bibinfo {author} {\bibfnamefont {M.}~\bibnamefont {Enderle}}, \bibinfo
    {author} {\bibfnamefont {Z.~Q.}\ \bibnamefont {Mao}}, \bibinfo {author}
    {\bibfnamefont {Y.}~\bibnamefont {Maeno}}, \bibinfo {author} {\bibfnamefont
    {I.~I.}\ \bibnamefont {Mazin}},\ and\ \bibinfo {author} {\bibfnamefont
    {S.~M.}\ \bibnamefont {Hayden}},\ }\bibfield  {title} {\enquote {\bibinfo
    {title} {Reduction of the Spin Susceptibility in the Superconducting State of
    ${\mathrm{Sr}}_{2}{\mathrm{RuO}}_{4}$ Observed by Polarized Neutron
    Scattering}}\ }\Doi {10.1103/PhysRevLett.125.217004} {\bibfield  {journal}
    {\bibinfo  {journal} {Phys. Rev. Lett.}\ }\textbf {\bibinfo {volume} {125}},\
    \bibinfo {pages} {217004} (\bibinfo {year} {2020})}\BibitemShut {NoStop}%
  \bibitem [{\citenamefont {Luke}\ \emph {et~al.}(1998)\citenamefont {Luke},
    \citenamefont {Fudamoto}, \citenamefont {Kojima}, \citenamefont {Larkin},
    \citenamefont {Merrin}, \citenamefont {Nachumi}, \citenamefont {Uemura},
    \citenamefont {Maeno}, \citenamefont {Mao}, \citenamefont {Mori},
    \citenamefont {Nakamura},\ and\ \citenamefont
    {Sigrist}}]{Luke1998.Nature.394.558}%
    \BibitemOpen
    \bibfield  {author} {\bibinfo {author} {\bibfnamefont {G.~M.}\ \bibnamefont
    {Luke}}, \bibinfo {author} {\bibfnamefont {Y.}~\bibnamefont {Fudamoto}},
    \bibinfo {author} {\bibfnamefont {K.~M.}\ \bibnamefont {Kojima}}, \bibinfo
    {author} {\bibfnamefont {M.~I.}\ \bibnamefont {Larkin}}, \bibinfo {author}
    {\bibfnamefont {J.}~\bibnamefont {Merrin}}, \bibinfo {author} {\bibfnamefont
    {B.}~\bibnamefont {Nachumi}}, \bibinfo {author} {\bibfnamefont {Y.~J.}\
    \bibnamefont {Uemura}}, \bibinfo {author} {\bibfnamefont {Y.}~\bibnamefont
    {Maeno}}, \bibinfo {author} {\bibfnamefont {Z.~Q.}\ \bibnamefont {Mao}},
    \bibinfo {author} {\bibfnamefont {Y.}~\bibnamefont {Mori}}, \bibinfo {author}
    {\bibfnamefont {H.}~\bibnamefont {Nakamura}},\ and\ \bibinfo {author}
    {\bibfnamefont {M.}~\bibnamefont {Sigrist}},\ }\bibfield  {title} {\enquote
    {\bibinfo {title} {Time-reversal symmetry-breaking superconductivity in
    Sr$_2$RuO$_4$}}\ }\Doi {10.1038/29038} {\bibfield  {journal} {\bibinfo
    {journal} {Nature}\ }\textbf {\bibinfo {volume} {394}},\ \bibinfo {pages}
    {558} (\bibinfo {year} {1998})}\BibitemShut {NoStop}%
  \bibitem [{\citenamefont {Xia}\ \emph {et~al.}(2006)\citenamefont {Xia},
    \citenamefont {Maeno}, \citenamefont {Beyersdorf}, \citenamefont {Fejer},\
    and\ \citenamefont {Kapitulnik}}]{Xia2006.PhysRevLett.97.167002}%
    \BibitemOpen
    \bibfield  {author} {\bibinfo {author} {\bibfnamefont {J.}~\bibnamefont
    {Xia}}, \bibinfo {author} {\bibfnamefont {Y.}~\bibnamefont {Maeno}}, \bibinfo
    {author} {\bibfnamefont {P.~T.}\ \bibnamefont {Beyersdorf}}, \bibinfo
    {author} {\bibfnamefont {M.~M.}\ \bibnamefont {Fejer}},\ and\ \bibinfo
    {author} {\bibfnamefont {A.}~\bibnamefont {Kapitulnik}},\ }\bibfield  {title}
    {\enquote {\bibinfo {title} {High Resolution Polar Kerr Effect Measurements
    of Sr$_2$RuO$_4$: Evidence for Broken Time-Reversal Symmetry in the
    Superconducting State}}\ }\Doi {10.1103/PhysRevLett.97.167002} {\bibfield
    {journal} {\bibinfo  {journal} {Phys. Rev. Lett.}\ }\textbf {\bibinfo
    {volume} {97}},\ \bibinfo {pages} {167002} (\bibinfo {year}
    {2006})}\BibitemShut {NoStop}%
  \bibitem [{\citenamefont {Grinenko}\ \emph
    {et~al.}(2021){\natexlab{a}}\citenamefont {Grinenko}, \citenamefont {Ghosh},
    \citenamefont {Sarkar}, \citenamefont {Orain}, \citenamefont {Nikitin},
    \citenamefont {Elender}, \citenamefont {Das}, \citenamefont {Guguchia},
    \citenamefont {Br^^c3^^bcckner}, \citenamefont {Barber}, \citenamefont
    {Park}, \citenamefont {Kikugawa}, \citenamefont {Sokolov}, \citenamefont
    {Bobowski}, \citenamefont {Miyoshi}, \citenamefont {Maeno}, \citenamefont
    {Mackenzie}, \citenamefont {Luetkens}, \citenamefont {Hicks},\ and\
    \citenamefont {Klauss}}]{Grinenko2021.Nat.Phys.17.748}%
    \BibitemOpen
    \bibfield  {author} {\bibinfo {author} {\bibfnamefont {V.}~\bibnamefont
    {Grinenko}}, \bibinfo {author} {\bibfnamefont {S.}~\bibnamefont {Ghosh}},
    \bibinfo {author} {\bibfnamefont {R.}~\bibnamefont {Sarkar}}, \bibinfo
    {author} {\bibfnamefont {J.-C.}\ \bibnamefont {Orain}}, \bibinfo {author}
    {\bibfnamefont {A.}~\bibnamefont {Nikitin}}, \bibinfo {author} {\bibfnamefont
    {M.}~\bibnamefont {Elender}}, \bibinfo {author} {\bibfnamefont
    {D.}~\bibnamefont {Das}}, \bibinfo {author} {\bibfnamefont {Z.}~\bibnamefont
    {Guguchia}}, \bibinfo {author} {\bibfnamefont {F.}~\bibnamefont
    {Br^^c3^^bcckner}}, \bibinfo {author} {\bibfnamefont {M.~E.}\ \bibnamefont
    {Barber}}, \bibinfo {author} {\bibfnamefont {J.}~\bibnamefont {Park}},
    \bibinfo {author} {\bibfnamefont {N.}~\bibnamefont {Kikugawa}}, \bibinfo
    {author} {\bibfnamefont {D.~A.}\ \bibnamefont {Sokolov}}, \bibinfo {author}
    {\bibfnamefont {J.~S.}\ \bibnamefont {Bobowski}}, \bibinfo {author}
    {\bibfnamefont {T.}~\bibnamefont {Miyoshi}}, \bibinfo {author} {\bibfnamefont
    {Y.}~\bibnamefont {Maeno}}, \bibinfo {author} {\bibfnamefont {A.~P.}\
    \bibnamefont {Mackenzie}}, \bibinfo {author} {\bibfnamefont {H.}~\bibnamefont
    {Luetkens}}, \bibinfo {author} {\bibfnamefont {C.~W.}\ \bibnamefont
    {Hicks}},\ and\ \bibinfo {author} {\bibfnamefont {H.-H.}\ \bibnamefont
    {Klauss}},\ }\bibfield  {title} {\enquote {\bibinfo {title} {Split
    superconducting and time-reversal symmetry-breaking transitions in
    Sr$_2$RuO$_4$ under stress}}\ }\Doi {10.1038/s41567-021-01182-7} {\bibfield
    {journal} {\bibinfo  {journal} {Nature Phys.}\ }\textbf {\bibinfo {volume}
    {17}},\ \bibinfo {pages} {748} (\bibinfo {year}
    {2021}{\natexlab{a}})}\BibitemShut {NoStop}%
  \bibitem [{\citenamefont {Grinenko}\ \emph
    {et~al.}(2021){\natexlab{b}}\citenamefont {Grinenko}, \citenamefont {Das},
    \citenamefont {Gupta}, \citenamefont {Zinkl}, \citenamefont {Kikugawa},
    \citenamefont {Maeno}, \citenamefont {Hicks}, \citenamefont {Klauss},
    \citenamefont {Sigrist},\ and\ \citenamefont
    {Khasanov}}]{Grinenko2021NatureComm.12.3920}%
    \BibitemOpen
    \bibfield  {author} {\bibinfo {author} {\bibfnamefont {V.}~\bibnamefont
    {Grinenko}}, \bibinfo {author} {\bibfnamefont {D.}~\bibnamefont {Das}},
    \bibinfo {author} {\bibfnamefont {R.}~\bibnamefont {Gupta}}, \bibinfo
    {author} {\bibfnamefont {B.}~\bibnamefont {Zinkl}}, \bibinfo {author}
    {\bibfnamefont {N.}~\bibnamefont {Kikugawa}}, \bibinfo {author}
    {\bibfnamefont {Y.}~\bibnamefont {Maeno}}, \bibinfo {author} {\bibfnamefont
    {C.~W.}\ \bibnamefont {Hicks}}, \bibinfo {author} {\bibfnamefont {H.-H.}\
    \bibnamefont {Klauss}}, \bibinfo {author} {\bibfnamefont {M.}~\bibnamefont
    {Sigrist}},\ and\ \bibinfo {author} {\bibfnamefont {R.}~\bibnamefont
    {Khasanov}},\ }\bibfield  {title} {\enquote {\bibinfo {title} {Unsplit
    superconducting and time reversal symmetry breaking transitions in Sr2RuO4
    under hydrostatic pressure and disorder}}\ }\Doi {10.1038/s41467-021-24176-8}
    {\bibfield  {journal} {\bibinfo  {journal} {Nature Communications}\ }\textbf
    {\bibinfo {volume} {12}},\ \bibinfo {pages} {3920} (\bibinfo {year}
    {2021}{\natexlab{b}})}\BibitemShut {NoStop}%
  \bibitem [{\citenamefont {Benhabib}\ \emph {et~al.}(2021)\citenamefont
    {Benhabib}, \citenamefont {Lupien}, \citenamefont {Paul}, \citenamefont
    {Berges}, \citenamefont {Dion}, \citenamefont {Nardone}, \citenamefont
    {Zitouni}, \citenamefont {Mao}, \citenamefont {Maeno}, \citenamefont
    {Georges}, \citenamefont {Taillefer},\ and\ \citenamefont
    {Proust}}]{Benhabib2021.Nat.Phys.17.194}%
    \BibitemOpen
    \bibfield  {author} {\bibinfo {author} {\bibfnamefont {S.}~\bibnamefont
    {Benhabib}}, \bibinfo {author} {\bibfnamefont {C.}~\bibnamefont {Lupien}},
    \bibinfo {author} {\bibfnamefont {I.}~\bibnamefont {Paul}}, \bibinfo {author}
    {\bibfnamefont {L.}~\bibnamefont {Berges}}, \bibinfo {author} {\bibfnamefont
    {M.}~\bibnamefont {Dion}}, \bibinfo {author} {\bibfnamefont {M.}~\bibnamefont
    {Nardone}}, \bibinfo {author} {\bibfnamefont {A.}~\bibnamefont {Zitouni}},
    \bibinfo {author} {\bibfnamefont {Z.~Q.}\ \bibnamefont {Mao}}, \bibinfo
    {author} {\bibfnamefont {Y.}~\bibnamefont {Maeno}}, \bibinfo {author}
    {\bibfnamefont {A.}~\bibnamefont {Georges}}, \bibinfo {author} {\bibfnamefont
    {L.}~\bibnamefont {Taillefer}},\ and\ \bibinfo {author} {\bibfnamefont
    {C.}~\bibnamefont {Proust}},\ }\bibfield  {title} {\enquote {\bibinfo {title}
    {Ultrasound evidence for a two-component superconducting order parameter in
    Sr$_2$RuO$_4$}}\ }\Doi {10.1038/s41567-020-1033-3} {\bibfield  {journal}
    {\bibinfo  {journal} {Nature Phys.}\ }\textbf {\bibinfo {volume} {17}},\
    \bibinfo {pages} {194} (\bibinfo {year} {2021})}\BibitemShut {NoStop}%
  \bibitem [{\citenamefont {Ghosh}\ \emph {et~al.}(2021)\citenamefont {Ghosh},
    \citenamefont {Shekhter}, \citenamefont {Jerzembeck}, \citenamefont
    {Kikugawa}, \citenamefont {Sokolov}, \citenamefont {Brando}, \citenamefont
    {Mackenzie}, \citenamefont {Hicks},\ and\ \citenamefont
    {Ramshaw}}]{Ghosh2021.Nat.Phys.17.199}%
    \BibitemOpen
    \bibfield  {author} {\bibinfo {author} {\bibfnamefont {S.}~\bibnamefont
    {Ghosh}}, \bibinfo {author} {\bibfnamefont {A.}~\bibnamefont {Shekhter}},
    \bibinfo {author} {\bibfnamefont {F.}~\bibnamefont {Jerzembeck}}, \bibinfo
    {author} {\bibfnamefont {N.}~\bibnamefont {Kikugawa}}, \bibinfo {author}
    {\bibfnamefont {D.~A.}\ \bibnamefont {Sokolov}}, \bibinfo {author}
    {\bibfnamefont {M.}~\bibnamefont {Brando}}, \bibinfo {author} {\bibfnamefont
    {A.~P.}\ \bibnamefont {Mackenzie}}, \bibinfo {author} {\bibfnamefont {C.~W.}\
    \bibnamefont {Hicks}},\ and\ \bibinfo {author} {\bibfnamefont {B.~J.}\
    \bibnamefont {Ramshaw}},\ }\bibfield  {title} {\enquote {\bibinfo {title}
    {Thermodynamic evidence for a two-component superconducting order parameter
    in Sr$_2$RuO$_4$}}\ }\Doi {10.1038/s41567-020-1032-4} {\bibfield  {journal}
    {\bibinfo  {journal} {Nature Phys.}\ }\textbf {\bibinfo {volume} {17}},\
    \bibinfo {pages} {199} (\bibinfo {year} {2021})}\BibitemShut {NoStop}%
  \bibitem [{\citenamefont {Ichioka}\ \emph {et~al.}(2007)\citenamefont
    {Ichioka}, \citenamefont {Adachi}, \citenamefont {Mizushima},\ and\
    \citenamefont {Machida}}]{Ichioka2007.PhysRevB.76.014503}%
    \BibitemOpen
    \bibfield  {author} {\bibinfo {author} {\bibfnamefont {M.}~\bibnamefont
    {Ichioka}}, \bibinfo {author} {\bibfnamefont {H.}~\bibnamefont {Adachi}},
    \bibinfo {author} {\bibfnamefont {T.}~\bibnamefont {Mizushima}},\ and\
    \bibinfo {author} {\bibfnamefont {K.}~\bibnamefont {Machida}},\ }\bibfield
    {title} {\enquote {\bibinfo {title} {Vortex state in a
    Fulde-Ferrell-Larkin-Ovchinnikov superconductor based on quasiclassical
    theory}}\ }\Doi {10.1103/PhysRevB.76.014503} {\bibfield  {journal} {\bibinfo
    {journal} {Phys. Rev. B}\ }\textbf {\bibinfo {volume} {76}},\ \bibinfo
    {pages} {014503} (\bibinfo {year} {2007})}\BibitemShut {NoStop}%
  \bibitem [{\citenamefont {Fulde}\ and\ \citenamefont
    {Ferrell}(1964)}]{Fulde1964PhysRev.135.A550}%
    \BibitemOpen
    \bibfield  {author} {\bibinfo {author} {\bibfnamefont {P.}~\bibnamefont
    {Fulde}}\ and\ \bibinfo {author} {\bibfnamefont {R.~A.}\ \bibnamefont
    {Ferrell}},\ }\bibfield  {title} {\enquote {\bibinfo {title}
    {Superconductivity in a Strong Spin-Exchange Field}}\ }\Doi
    {10.1103/PhysRev.135.A550} {\bibfield  {journal} {\bibinfo  {journal} {Phys.
    Rev.}\ }\textbf {\bibinfo {volume} {135}},\ \bibinfo {pages} {A550} (\bibinfo
    {year} {1964})}\BibitemShut {NoStop}%
  \bibitem [{\citenamefont {Larkin}\ and\ \citenamefont
    {Ovchinnikov}(1964)}]{Larkin1964ZhEksp}%
    \BibitemOpen
    \bibfield  {author} {\bibinfo {author} {\bibfnamefont {A.~I.}\ \bibnamefont
    {Larkin}}\ and\ \bibinfo {author} {\bibfnamefont {Y.~N.}\ \bibnamefont
    {Ovchinnikov}},\ }\href@noop {} {\bibfield  {journal} {\bibinfo  {journal}
    {Zh. Eksp. Teor. Fiz.}\ }\textbf {\bibinfo {volume} {47}},\ \bibinfo {pages}
    {1136} (\bibinfo {year} {1964})},\ \bibinfo {note} {[translation: Sov. Phys.
    JETP {\bf 20} (1965) 762]}\BibitemShut {NoStop}%
  \bibitem [{\citenamefont {Yonezawa}\ \emph {et~al.}(2013)\citenamefont
    {Yonezawa}, \citenamefont {Kajikawa},\ and\ \citenamefont
    {Maeno}}]{Yonezawa2013.PhysRevLett.110.077003}%
    \BibitemOpen
    \bibfield  {author} {\bibinfo {author} {\bibfnamefont {S.}~\bibnamefont
    {Yonezawa}}, \bibinfo {author} {\bibfnamefont {T.}~\bibnamefont {Kajikawa}},\
    and\ \bibinfo {author} {\bibfnamefont {Y.}~\bibnamefont {Maeno}},\ }\bibfield
     {title} {\enquote {\bibinfo {title} {First-Order Superconducting Transition
    of Sr$_2$RuO$_4$}}\ }\Doi {10.1103/PhysRevLett.110.077003} {\bibfield
    {journal} {\bibinfo  {journal} {Phys. Rev. Lett.}\ }\textbf {\bibinfo
    {volume} {110}},\ \bibinfo {pages} {077003} (\bibinfo {year}
    {2013})}\BibitemShut {NoStop}%
  \bibitem [{\citenamefont {Yonezawa}\ \emph {et~al.}(2014)\citenamefont
    {Yonezawa}, \citenamefont {Kajikawa},\ and\ \citenamefont
    {Maeno}}]{Yonezawa2014.JPhysSocJpn.83.083706}%
    \BibitemOpen
    \bibfield  {author} {\bibinfo {author} {\bibfnamefont {S.}~\bibnamefont
    {Yonezawa}}, \bibinfo {author} {\bibfnamefont {T.}~\bibnamefont {Kajikawa}},\
    and\ \bibinfo {author} {\bibfnamefont {Y.}~\bibnamefont {Maeno}},\ }\bibfield
     {title} {\enquote {\bibinfo {title} {Specific-Heat Evidence of the
    First-Order Superconducting Transition in Sr$_2$RuO$_4$}}\ }\Doi
    {10.7566/JPSJ.83.083706} {\bibfield  {journal} {\bibinfo  {journal} {J. Phys.
    Soc. Jpn.}\ }\textbf {\bibinfo {volume} {83}},\ \bibinfo {pages} {083706}
    (\bibinfo {year} {2014})}\BibitemShut {NoStop}%
  \bibitem [{\citenamefont {Mao}\ \emph {et~al.}(2000){\natexlab{a}}\citenamefont
    {Mao}, \citenamefont {Maeno}, \citenamefont {NishiZaki}, \citenamefont
    {Akima},\ and\ \citenamefont {Ishiguro}}]{Mao2000.PhysRevLett.84.991}%
    \BibitemOpen
    \bibfield  {author} {\bibinfo {author} {\bibfnamefont {Z.~Q.}\ \bibnamefont
    {Mao}}, \bibinfo {author} {\bibfnamefont {Y.}~\bibnamefont {Maeno}}, \bibinfo
    {author} {\bibfnamefont {S.}~\bibnamefont {NishiZaki}}, \bibinfo {author}
    {\bibfnamefont {T.}~\bibnamefont {Akima}},\ and\ \bibinfo {author}
    {\bibfnamefont {T.}~\bibnamefont {Ishiguro}},\ }\bibfield  {title} {\enquote
    {\bibinfo {title} {In-Plane Anisotropy of Upper Critical Field in
    Sr$_2$RuO$_4$}}\ }\Doi {10.1103/PhysRevLett.84.991} {\bibfield  {journal}
    {\bibinfo  {journal} {Phys. Rev. Lett.}\ }\textbf {\bibinfo {volume} {84}},\
    \bibinfo {pages} {991} (\bibinfo {year} {2000}{\natexlab{a}})}\BibitemShut
    {NoStop}%
  \bibitem [{\citenamefont {Kittaka}\ \emph {et~al.}(2009)\citenamefont
    {Kittaka}, \citenamefont {Nakamura}, \citenamefont {Aono}, \citenamefont
    {Yonezawa}, \citenamefont {Ishida},\ and\ \citenamefont
    {Maeno}}]{Kittaka2009.PhysRevB.80.174514}%
    \BibitemOpen
    \bibfield  {author} {\bibinfo {author} {\bibfnamefont {S.}~\bibnamefont
    {Kittaka}}, \bibinfo {author} {\bibfnamefont {T.}~\bibnamefont {Nakamura}},
    \bibinfo {author} {\bibfnamefont {Y.}~\bibnamefont {Aono}}, \bibinfo {author}
    {\bibfnamefont {S.}~\bibnamefont {Yonezawa}}, \bibinfo {author}
    {\bibfnamefont {K.}~\bibnamefont {Ishida}},\ and\ \bibinfo {author}
    {\bibfnamefont {Y.}~\bibnamefont {Maeno}},\ }\bibfield  {title} {\enquote
    {\bibinfo {title} {Angular dependence of the upper critical field of
    Sr$_2$RuO$_4$}}\ }\Doi {10.1103/PhysRevB.80.174514} {\bibfield  {journal}
    {\bibinfo  {journal} {Phys. Rev. B}\ }\textbf {\bibinfo {volume} {80}},\
    \bibinfo {pages} {174514} (\bibinfo {year} {2009})}\BibitemShut {NoStop}%
  \bibitem [{\citenamefont {Matsushita}(2014)}]{MatsushitaText}%
    \BibitemOpen
    \bibfield  {author} {\bibinfo {author} {\bibfnamefont {T.}~\bibnamefont
    {Matsushita}},\ }\href@noop {} {\emph {\bibinfo {title} {Flux pinning in
    superconductors}}}\ (\bibinfo  {publisher} {Springer},\ \bibinfo {year}
    {2014})\BibitemShut {NoStop}%
  \bibitem [{\citenamefont {Mackenzie}\ \emph {et~al.}(1998)\citenamefont
    {Mackenzie}, \citenamefont {Haselwimmer}, \citenamefont {Tyler},
    \citenamefont {Lonzarich}, \citenamefont {Mori}, \citenamefont {Nishizaki},\
    and\ \citenamefont {Maeno}}]{Mackenzie1998.PhysRevLett.80.161}%
    \BibitemOpen
    \bibfield  {author} {\bibinfo {author} {\bibfnamefont {A.~P.}\ \bibnamefont
    {Mackenzie}}, \bibinfo {author} {\bibfnamefont {R.~K.~W.}\ \bibnamefont
    {Haselwimmer}}, \bibinfo {author} {\bibfnamefont {A.~W.}\ \bibnamefont
    {Tyler}}, \bibinfo {author} {\bibfnamefont {G.~G.}\ \bibnamefont
    {Lonzarich}}, \bibinfo {author} {\bibfnamefont {Y.}~\bibnamefont {Mori}},
    \bibinfo {author} {\bibfnamefont {S.}~\bibnamefont {Nishizaki}},\ and\
    \bibinfo {author} {\bibfnamefont {Y.}~\bibnamefont {Maeno}},\ }\bibfield
    {title} {\enquote {\bibinfo {title} {Extremely Strong Dependence of
    Superconductivity on Disorder in Sr$_2$RuO$_4$}}\ }\Doi
    {10.1103/PhysRevLett.80.161} {\bibfield  {journal} {\bibinfo  {journal}
    {Phys. Rev. Lett.}\ }\textbf {\bibinfo {volume} {80}},\ \bibinfo {pages}
    {161} (\bibinfo {year} {1998})}\BibitemShut {NoStop}%
  \bibitem [{\citenamefont {Mao}\ \emph {et~al.}(2000){\natexlab{b}}\citenamefont
    {Mao}, \citenamefont {Maeno},\ and\ \citenamefont
    {Fukazawa}}]{Mao2000.MaterResBull.35.1813}%
    \BibitemOpen
    \bibfield  {author} {\bibinfo {author} {\bibfnamefont {Z.}~\bibnamefont
    {Mao}}, \bibinfo {author} {\bibfnamefont {Y.}~\bibnamefont {Maeno}},\ and\
    \bibinfo {author} {\bibfnamefont {H.}~\bibnamefont {Fukazawa}},\ }\bibfield
    {title} {\enquote {\bibinfo {title} {Crystal growth of Sr$_2$RuO$_4$}}\ }\Doi
    {10.1016/S0025-5408(00)00378-0} {\bibfield  {journal} {\bibinfo  {journal}
    {Mater. Res. Bull.}\ }\textbf {\bibinfo {volume} {35}},\ \bibinfo {pages}
    {1813} (\bibinfo {year} {2000}{\natexlab{b}})}\BibitemShut {NoStop}%
  \bibitem [{Sup()}]{SupplementalAraki}%
    \BibitemOpen
    \href@noop {} {\bibinfo  {journal} {See Supplemental Material at [URL will be
    inserted by publisher] for [give brief description of
    material].}}\BibitemShut {Stop}%
  \bibitem [{\citenamefont {Yonezawa}\ \emph {et~al.}(2015)\citenamefont
    {Yonezawa}, \citenamefont {Higuchi}, \citenamefont {Sugimoto}, \citenamefont
    {Sow},\ and\ \citenamefont {Maeno}}]{Yonezawa2015.RevSciInstrum.86.093903}%
    \BibitemOpen
  \bibfield  {journal} {  }\bibfield  {author} {\bibinfo {author} {\bibfnamefont
    {S.}~\bibnamefont {Yonezawa}}, \bibinfo {author} {\bibfnamefont
    {T.}~\bibnamefont {Higuchi}}, \bibinfo {author} {\bibfnamefont
    {Y.}~\bibnamefont {Sugimoto}}, \bibinfo {author} {\bibfnamefont
    {C.}~\bibnamefont {Sow}},\ and\ \bibinfo {author} {\bibfnamefont
    {Y.}~\bibnamefont {Maeno}},\ }\bibfield  {title} {\enquote {\bibinfo {title}
    {Compact AC susceptometer for fast sample characterization down to 0.1 K}}\
    }\Doi {10.1063/1.4929871} {\bibfield  {journal} {\bibinfo  {journal} {Rev.
    Sci. Intrum.}\ }\textbf {\bibinfo {volume} {86}},\ \bibinfo {pages} {093903}
    (\bibinfo {year} {2015})}\BibitemShut {NoStop}%
  \bibitem [{\citenamefont {Loetzsch}\ \emph {et~al.}(2010)\citenamefont
    {Loetzsch}, \citenamefont {L^^c3^^bcbcke}, \citenamefont {Uschmann},
    \citenamefont {F^^c3^^b6rster}, \citenamefont {Gro^^c3^^9fe}, \citenamefont
    {Thuerk}, \citenamefont {Koettig}, \citenamefont {Schmidl},\ and\
    \citenamefont {Seidel}}]{Loetzsch2010APL.96.071901}%
    \BibitemOpen
    \bibfield  {author} {\bibinfo {author} {\bibfnamefont {R.}~\bibnamefont
    {Loetzsch}}, \bibinfo {author} {\bibfnamefont {A.}~\bibnamefont
    {L^^c3^^bcbcke}}, \bibinfo {author} {\bibfnamefont {I.}~\bibnamefont
    {Uschmann}}, \bibinfo {author} {\bibfnamefont {E.}~\bibnamefont
    {F^^c3^^b6rster}}, \bibinfo {author} {\bibfnamefont {V.}~\bibnamefont
    {Gro^^c3^^9fe}}, \bibinfo {author} {\bibfnamefont {M.}~\bibnamefont
    {Thuerk}}, \bibinfo {author} {\bibfnamefont {T.}~\bibnamefont {Koettig}},
    \bibinfo {author} {\bibfnamefont {F.}~\bibnamefont {Schmidl}},\ and\ \bibinfo
    {author} {\bibfnamefont {P.}~\bibnamefont {Seidel}},\ }\bibfield  {title}
    {\enquote {\bibinfo {title} {The cubic to tetragonal phase transition in
    SrTiO$_3$ single crystals near its surface under internal and external
    strains}}\ }\Doi {10.1063/1.3324695} {\bibfield  {journal} {\bibinfo
    {journal} {Appl. Phys. Lett.}\ }\textbf {\bibinfo {volume} {96}},\ \bibinfo
    {pages} {071901} (\bibinfo {year} {2010})}\BibitemShut {NoStop}%
  \bibitem [{\citenamefont {Chmaissem}\ \emph {et~al.}(1998)\citenamefont
    {Chmaissem}, \citenamefont {Jorgensen}, \citenamefont {Shaked}, \citenamefont
    {Ikeda},\ and\ \citenamefont {Maeno}}]{Chmaissem1998PhysRevB.57.5067}%
    \BibitemOpen
    \bibfield  {author} {\bibinfo {author} {\bibfnamefont {O.}~\bibnamefont
    {Chmaissem}}, \bibinfo {author} {\bibfnamefont {J.~D.}\ \bibnamefont
    {Jorgensen}}, \bibinfo {author} {\bibfnamefont {H.}~\bibnamefont {Shaked}},
    \bibinfo {author} {\bibfnamefont {S.}~\bibnamefont {Ikeda}},\ and\ \bibinfo
    {author} {\bibfnamefont {Y.}~\bibnamefont {Maeno}},\ }\bibfield  {title}
    {\enquote {\bibinfo {title} {Thermal expansion and compressibility of
    ${\mathrm{Sr}}_{2}{\mathrm{RuO}}_{4}$}}\ }\Doi {10.1103/PhysRevB.57.5067}
    {\bibfield  {journal} {\bibinfo  {journal} {Phys. Rev. B}\ }\textbf {\bibinfo
    {volume} {57}},\ \bibinfo {pages} {5067} (\bibinfo {year}
    {1998})}\BibitemShut {NoStop}%
  \bibitem [{\citenamefont {Yasui}\ \emph {et~al.}(2017)\citenamefont {Yasui},
    \citenamefont {Lahabi}, \citenamefont {Anwar}, \citenamefont {Nakamura},
    \citenamefont {Yonezawa}, \citenamefont {Terashima}, \citenamefont {Aarts},\
    and\ \citenamefont {Maeno}}]{Yasui2017PhysRevB.96.180507}%
    \BibitemOpen
    \bibfield  {author} {\bibinfo {author} {\bibfnamefont {Y.}~\bibnamefont
    {Yasui}}, \bibinfo {author} {\bibfnamefont {K.}~\bibnamefont {Lahabi}},
    \bibinfo {author} {\bibfnamefont {M.~S.}\ \bibnamefont {Anwar}}, \bibinfo
    {author} {\bibfnamefont {Y.}~\bibnamefont {Nakamura}}, \bibinfo {author}
    {\bibfnamefont {S.}~\bibnamefont {Yonezawa}}, \bibinfo {author}
    {\bibfnamefont {T.}~\bibnamefont {Terashima}}, \bibinfo {author}
    {\bibfnamefont {J.}~\bibnamefont {Aarts}},\ and\ \bibinfo {author}
    {\bibfnamefont {Y.}~\bibnamefont {Maeno}},\ }\bibfield  {title} {\enquote
    {\bibinfo {title} {Little-Parks oscillations with half-quantum fluxoid
    features in ${\mathrm{Sr}}_{2}{\mathrm{RuO}}_{4}$ microrings}}\ }\Doi
    {10.1103/PhysRevB.96.180507} {\bibfield  {journal} {\bibinfo  {journal}
    {Phys. Rev. B}\ }\textbf {\bibinfo {volume} {96}},\ \bibinfo {pages} {180507}
    (\bibinfo {year} {2017})}\BibitemShut {NoStop}%
  \bibitem [{\citenamefont {Deguchi}\ \emph {et~al.}(2004)\citenamefont
    {Deguchi}, \citenamefont {Ishiguro},\ and\ \citenamefont
    {Maeno}}]{Deguchi2004RSI}%
    \BibitemOpen
    \bibfield  {author} {\bibinfo {author} {\bibfnamefont {K.}~\bibnamefont
    {Deguchi}}, \bibinfo {author} {\bibfnamefont {T.}~\bibnamefont {Ishiguro}},\
    and\ \bibinfo {author} {\bibfnamefont {Y.}~\bibnamefont {Maeno}},\ }\bibfield
     {title} {\enquote {\bibinfo {title} {Field-orientation dependent heat
    capacity measurements at low temperatures with a vector magnet system}}\
    }\href@noop {} {\bibfield  {journal} {\bibinfo  {journal} {Rev. Sci.
    Instrum.}\ }\textbf {\bibinfo {volume} {75}},\ \bibinfo {pages} {1188}
    (\bibinfo {year} {2004})}\BibitemShut {NoStop}%
  \bibitem [{\citenamefont {Kinjo}\ \emph {et~al.}(2021)\citenamefont {Kinjo},
    \citenamefont {Manago}, \citenamefont {Kitagawa}, \citenamefont {Mao},
    \citenamefont {Yonezawa}, \citenamefont {Maeno},\ and\ \citenamefont
    {Ishida}}]{Kinjoimplement}%
    \BibitemOpen
    \bibfield  {author} {\bibinfo {author} {\bibfnamefont {K.}~\bibnamefont
    {Kinjo}}, \bibinfo {author} {\bibfnamefont {M.}~\bibnamefont {Manago}},
    \bibinfo {author} {\bibfnamefont {S.}~\bibnamefont {Kitagawa}}, \bibinfo
    {author} {\bibfnamefont {Z.~Q.}\ \bibnamefont {Mao}}, \bibinfo {author}
    {\bibfnamefont {S.}~\bibnamefont {Yonezawa}}, \bibinfo {author}
    {\bibfnamefont {Y.}~\bibnamefont {Maeno}},\ and\ \bibinfo {author}
    {\bibfnamefont {K.}~\bibnamefont {Ishida}},\ }\bibfield  {title} {\enquote
    {\bibinfo {title} {Superconducting Spin Smecticity Evidencing the
    Fulde-Ferrell-Larkin-Ovchinnikov State in Sr$_2$RuO$_4$}}\ }\href@noop {}
    {\bibfield  {journal} {\bibinfo  {journal} {submitted}} (\bibinfo {year}
    {2021})}\BibitemShut {NoStop}%
  \bibitem [{\citenamefont {Yonezawa}(2019)}]{Yonezawa2019condmat4010002}%
    \BibitemOpen
    \bibfield  {author} {\bibinfo {author} {\bibfnamefont {S.}~\bibnamefont
    {Yonezawa}},\ }\bibfield  {title} {\enquote {\bibinfo {title} {Nematic
    Superconductivity in Doped Bi2Se3 Topological Superconductors}}\ }\Doi
    {10.3390/condmat4010002} {\bibfield  {journal} {\bibinfo  {journal}
    {Condensed Matter}\ }\textbf {\bibinfo {volume} {4}} (\bibinfo {year}
    {2019})},\ ISSN \bibinfo {issn} {2410-3896},\ \doi
    {10.3390/condmat4010002}\BibitemShut {NoStop}%
  \bibitem [{\citenamefont {Fernandes}\ \emph {et~al.}(2014)\citenamefont
    {Fernandes}, \citenamefont {Chubukov},\ and\ \citenamefont
    {Schmalian}}]{Fernandes2014nematicorder}%
    \BibitemOpen
    \bibfield  {author} {\bibinfo {author} {\bibfnamefont {R.~M.}\ \bibnamefont
    {Fernandes}}, \bibinfo {author} {\bibfnamefont {A.~V.}\ \bibnamefont
    {Chubukov}},\ and\ \bibinfo {author} {\bibfnamefont {J.}~\bibnamefont
    {Schmalian}},\ }\bibfield  {title} {\enquote {\bibinfo {title} {What drives
    nematic order in iron-based^^c2^^a0superconductors?}}\ }\Doi
    {10.1038/nphys2877} {\bibfield  {journal} {\bibinfo  {journal} {Nature
    Physics}\ }\textbf {\bibinfo {volume} {10}},\ \bibinfo {pages} {97} (\bibinfo
    {year} {2014})},\ ISSN \bibinfo {issn} {1745-2481}\BibitemShut {NoStop}%
  \end{thebibliography}
%


\clearpage

 \onecolumngrid
 \begin{center}
   \textbf{\large Supplemental Material for\\
   Current-induced superconducting anisotropy of Sr$\mathsf{_2}$RuO$\mathsf{_4}$}\\[.2cm]
   R.~Araki,$^{1,*}$ T.~Miyoshi,$^{1}$ H.~Suwa,${^1}$ E.~I.~Paredes Aulestia,$^{2}$ K.~Y.~Yip,$^{2}$
    K.~T.~Lai,$^{2}$ S.~K.~Goh,$^{2}$ Y.~Maeno,$^{1}$ and S.~Yonezawa$^{1,\dagger}$\\[.1cm]
   {\it ${}^1$Department of Physics, Kyoto University, Kyoto 606-8502, Japan\\
   ${}^2$Department of Physics, The Chinese University of Hong Kong, Shatin N.T., Hong Kong}\\
   ${}^*$araki.ryo.43w@st.kyoto-u.ac.jp\\
   ${}^{\dagger}$yonezawa.shingo.3m@kyoto-u.ac.jp\\
 (Dated: \today)\\[1cm]
 \end{center}
 
 \setcounter{equation}{0}
 \setcounter{figure}{0}
 \setcounter{table}{0}
 \setcounter{page}{1}
 \renewcommand{\theequation}{S\arabic{equation}}
 \renewcommand{\thefigure}{S\arabic{figure}}
 \renewcommand{\bibnumfmt}[1]{[S#1]}
 \renewcommand{\citenumfont}[1]{S#1}

 \begin{figure}[htb]
   \centering
   \includegraphics[width=12.0cm]{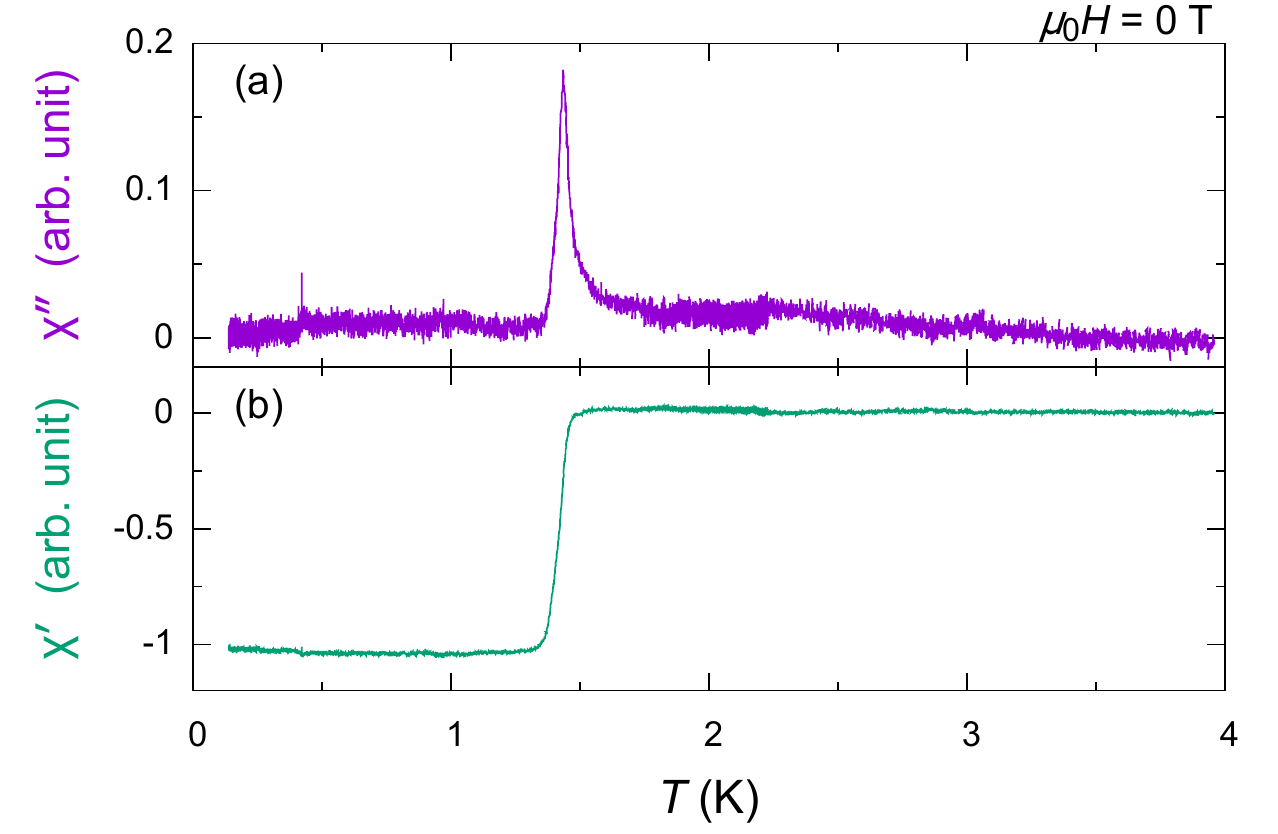}
   \caption{AC susceptibility $\chi_{\soeji{ac}} \equiv \chi^{\prime} -i \chi^{\prime\prime}$ of the ${\rm \Large Sr_2RuO_4}$ single crystal used for our device. 
   Before the micro-fabrication, 
    ac susceptibility measurement at zero DC field revealed that this thin crystal (batch: C432; 500~$\muup$m $\times$
    100~$\muup$m $\times$ 10~$\muup$m) exhibits superconducting transition at $T_{\soeji{c}} = 1.43~$K,
    which is defined as the peak of the imaginary part of $\chi_{\soeji{ac}}$.
   }\label{kakoumae}
 \end{figure}
 

  

  \begin{figure}[htb]
    \centering
    \includegraphics[width=6.5cm]{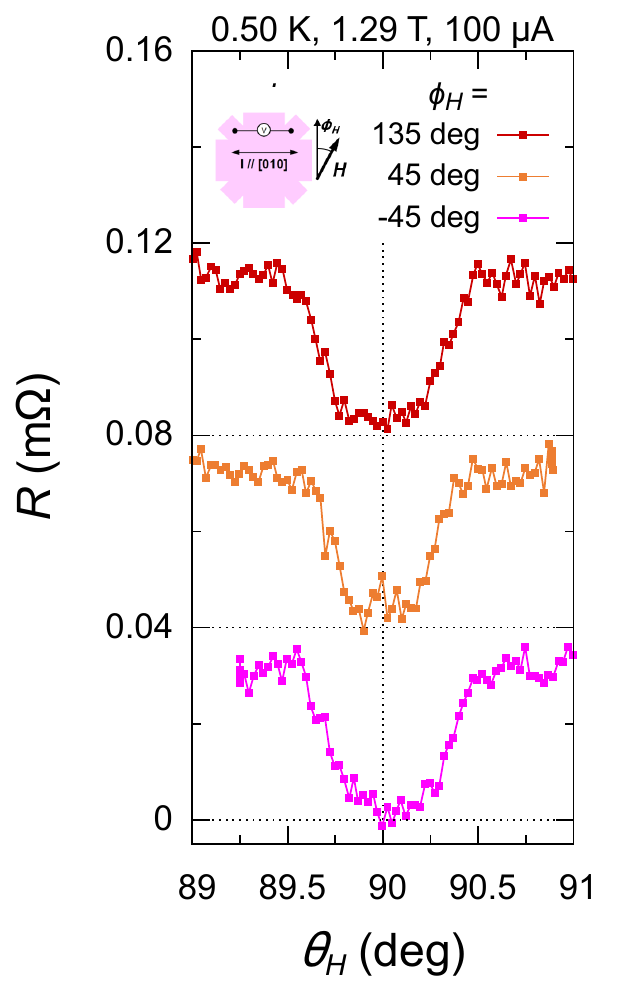}
    \caption{Check of the alignment of magnetic field to the $ab$ plane. 
    The figure shows the polar-angle $\theta_{H}$ dependence of the resistance $R$ measured at 0.50~K under 1.29~T and 100-$\muup$A
    current along the [010] direction. The curves have vertical offsets.
    The red, orange and pink curves correspond to $\phi_H = 135^{\circ}, 45^{\circ}$ and $-45^{\circ}$, respectively.
    For all $\phi_H$, $R(\theta_H)$ curves show sharp and symmetric dips centered at $\theta_H = 90^{\circ}$ due to superconductivity.
     These data indicate that the misalignment of the magnetic field with respect to the $ab$ plane $(\theta_H = 90^{\circ})$ is at most $0.1^{\circ}$.
    }\label{misalignment}
    \end{figure}

    

    \begin{figure*}[htb]
      \begin{center}
        \def\subfigcapskip{-3pt}
       \subfigure{
        \includegraphics[width=.45\columnwidth]{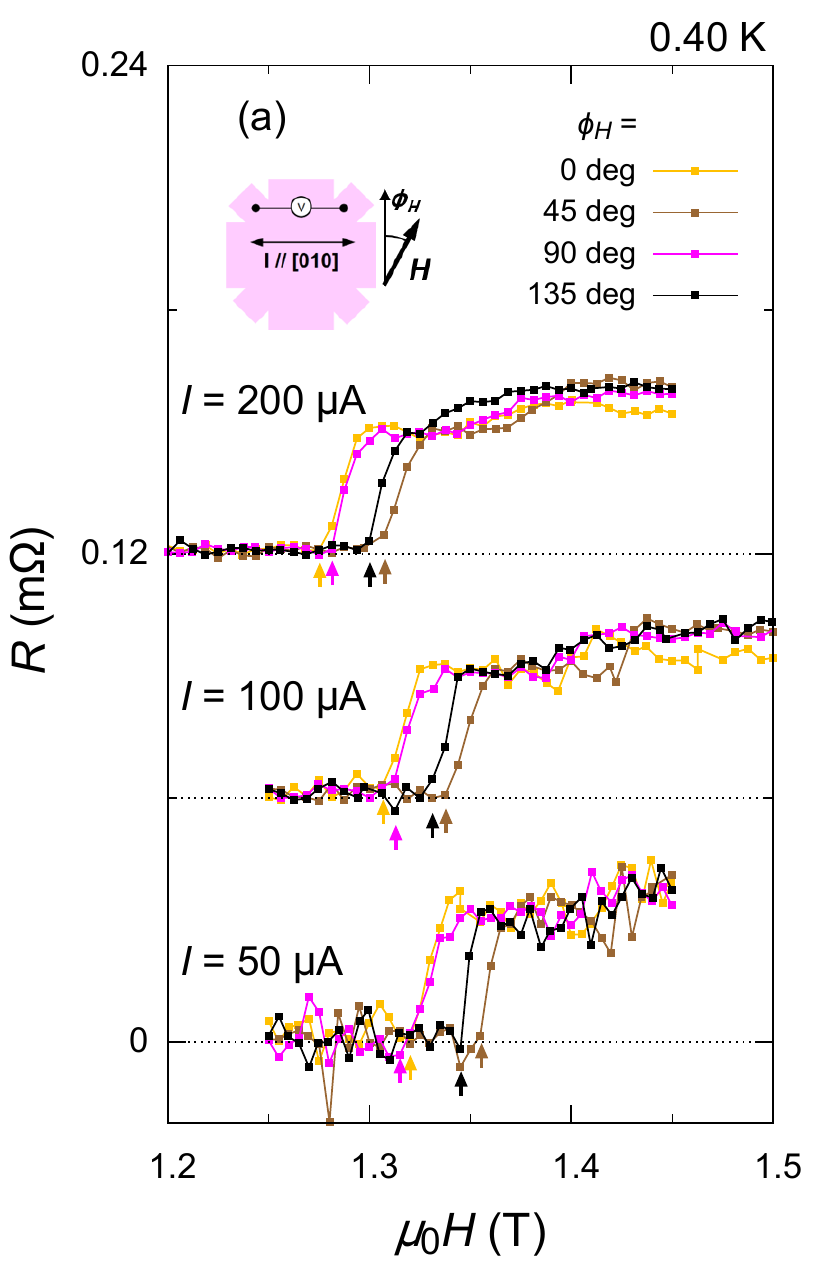} 
       }
       \def\subfigcapskip{-3pt}
       \subfigure{
        \includegraphics[width=.45\columnwidth]{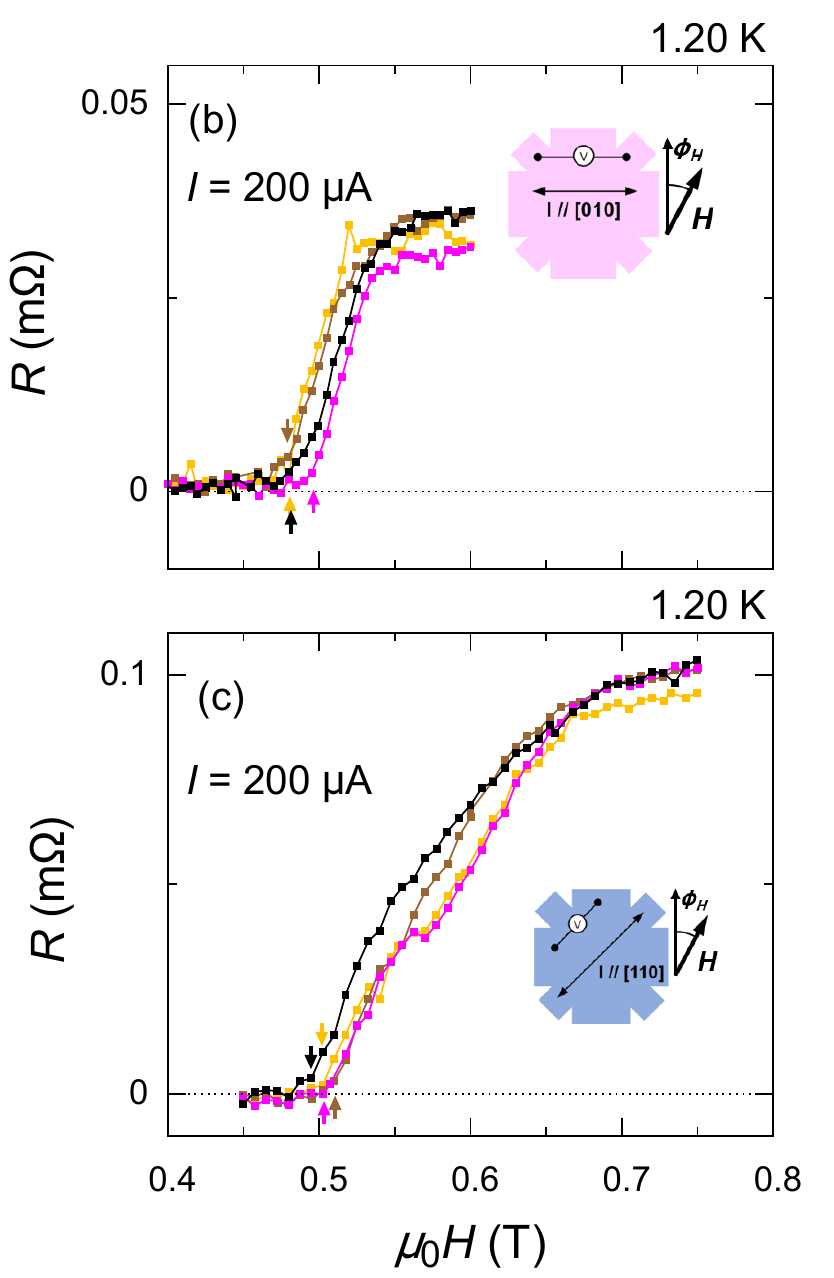}
       }
      \end{center}
      \caption{Raw magneto-resistance data. 
      The figure represents resistive transition of our ${\rm Sr_2RuO_4}$ device under currents.
      The yellow, brown, pink and black data points are obtained under $\phi_H = 0^{\circ}$, $45^{\circ}$, $90^{\circ}$ and $135^{\circ}$
      , respectively.
      (a) Data under currents along the [010] direction at 0.4~K.
      Similarly to the $R(H)$ curves in Fig.~2 of the main text, $H_{\soeji{c2}}$, indicated by the vertical arrows,
       exhibits two-fold anisotropy as evidenced by the differences between the curves 
       of $\phi_H = 0^{\circ}$ and $90^{\circ}$, or those of $\phi_H = 45^{\circ}$
       and $135^{\circ}$. 
      (b) Data under currents along the [010] direction at 1.20~K, where superconducting transition is an ordinary
       second-order transition.
      (c) Data under currents along the [110] direction at 1.20~K.
      }
    \end{figure*}

    \begin{figure*}[htb]
      \begin{center}
        \def\subfigcapskip{-3pt}
       \subfigure{
        \includegraphics[width=.32\columnwidth]{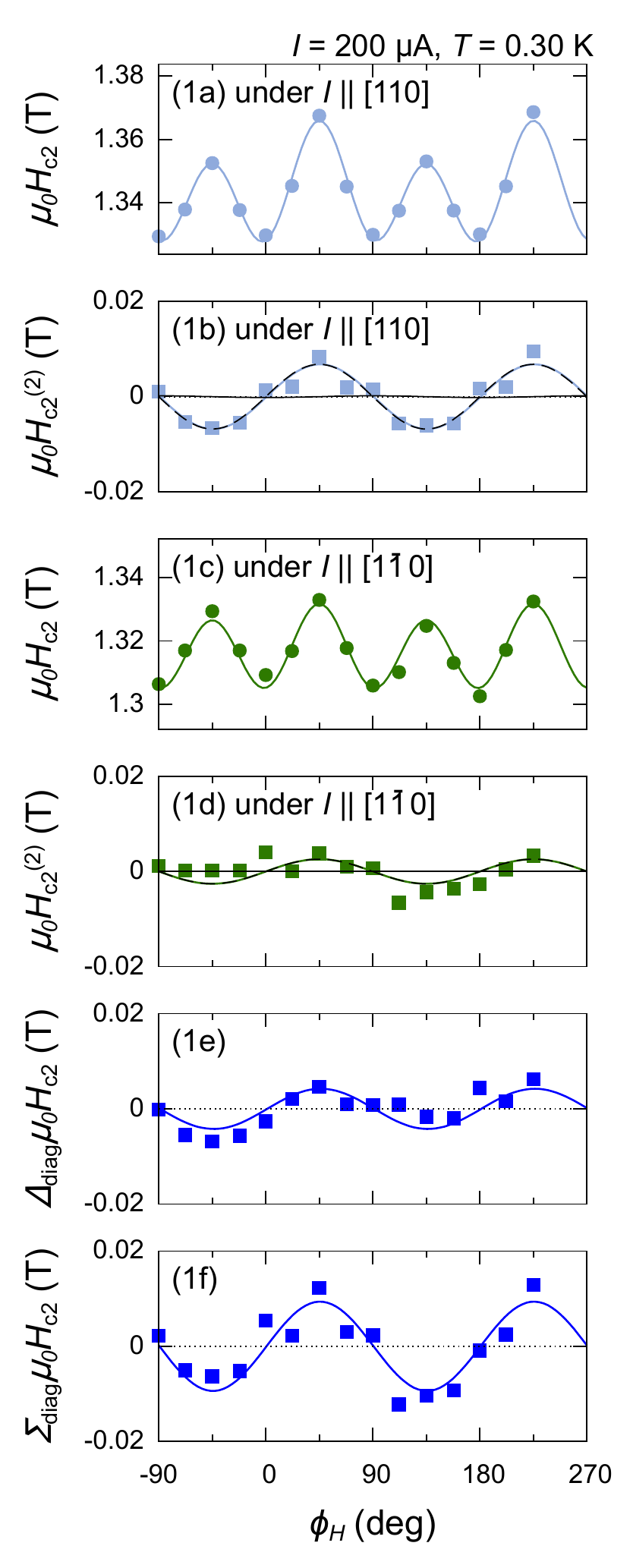} 
       }
       \def\subfigcapskip{-3pt}
       \subfigure{
        \includegraphics[width=.32\columnwidth]{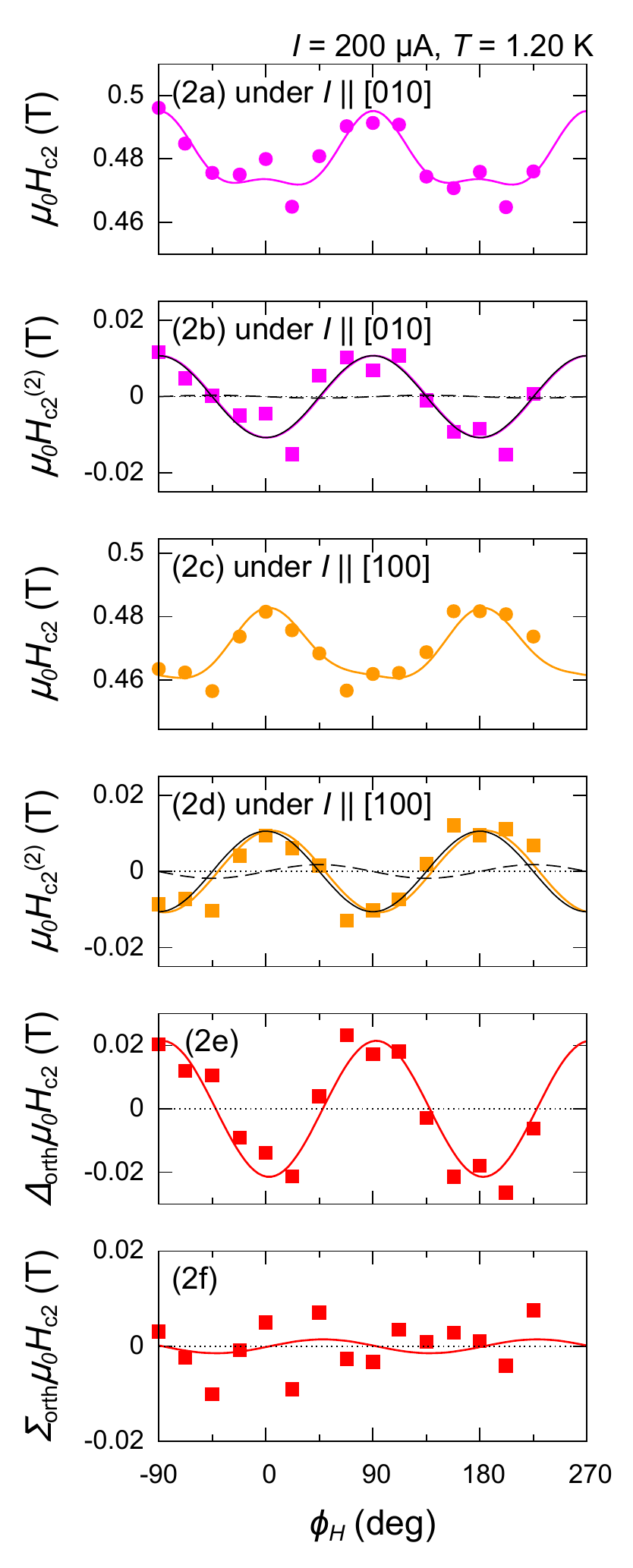}
       }
       \def\subfigcapskip{-3pt}
       \subfigure{
        \includegraphics[width=.32\columnwidth]{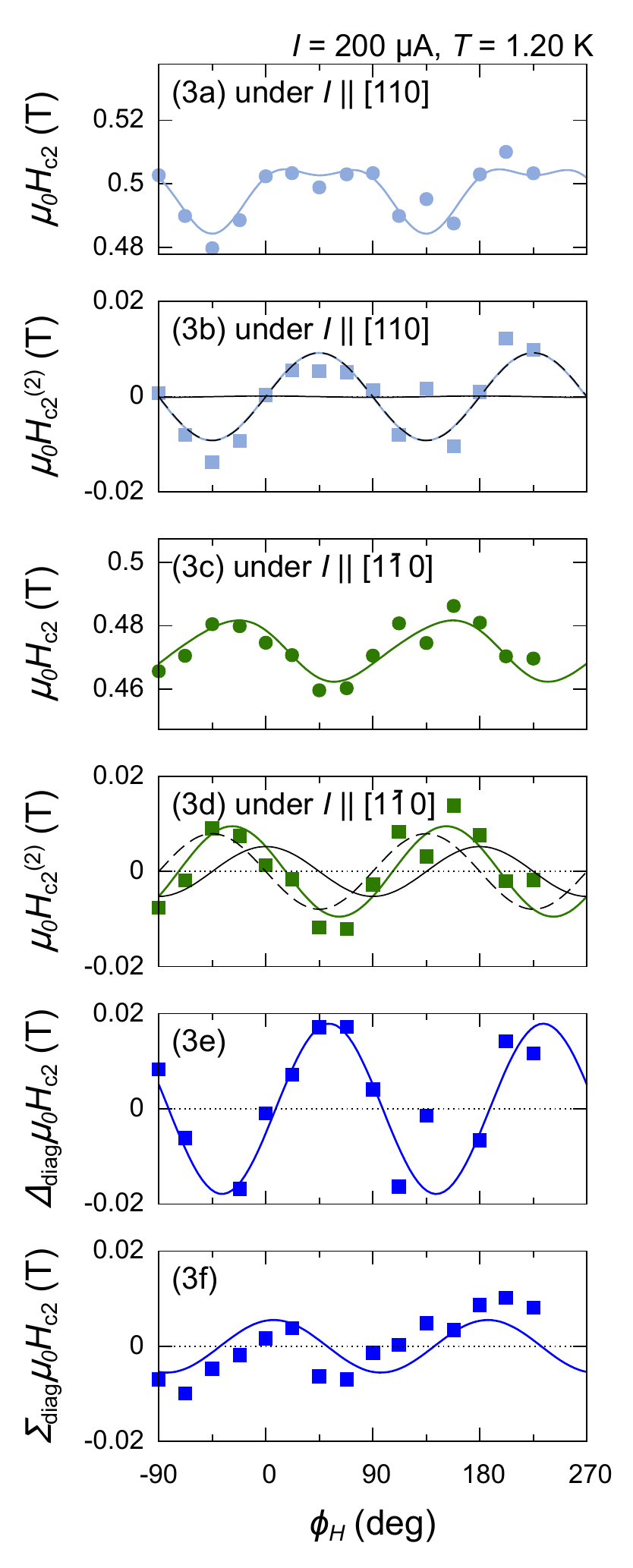} 
       }
      \end{center}
      \caption{Representative $H_{\soeji{c2}}$ anisotropy of ${\rm Sr_2 Ru O_4}$ under in-plane current.
      Similarly to Fig.~3 of the main text, (a) and (c) represent raw $H_{\soeji{c2}}$ data (circles)
       and fitting result (solid curve) under current, (b) and (d) represent two-fold components of $H_{\soeji{c2}}$ under current,
       extracted from the raw $H_{\soeji{c2}}$ data by subtracting the fitted four-fold component and the constant offset,
       (e) represents $I$-induced two-fold component of $H_{\soeji{c2}}$ obtained by
       taking the difference between the data in (b) and (d), and
      (f) represents $I$-independent two-fold component of $H_{\soeji{c2}}$ obtained by
       taking the summation of the data in (b) and (d), respectively.
      The sky blue, green, pink and yellow color indicate data under current parallel to the [110], $[1\overline{1}0]$,
       [010] and [100] directions, respectively. 
      The blue and red color indicate data evaluated from the datasets for the orthogonal and diagonal current-directions, respectively.    
      Column (1) $H_{\soeji{c2}}$ under currents along the diagonal directions at 0.30~K. 
      Column (2) $H_{\soeji{c2}}$ under currents along the orthogonal directions at 1.20~K.
      Column (3) $H_{\soeji{c2}}$ under currents along the diagonal directions at 1.20~K.
      }
    \end{figure*}

    
    \begin{figure}[htb]
      \centering
      \includegraphics[width=12.0cm]{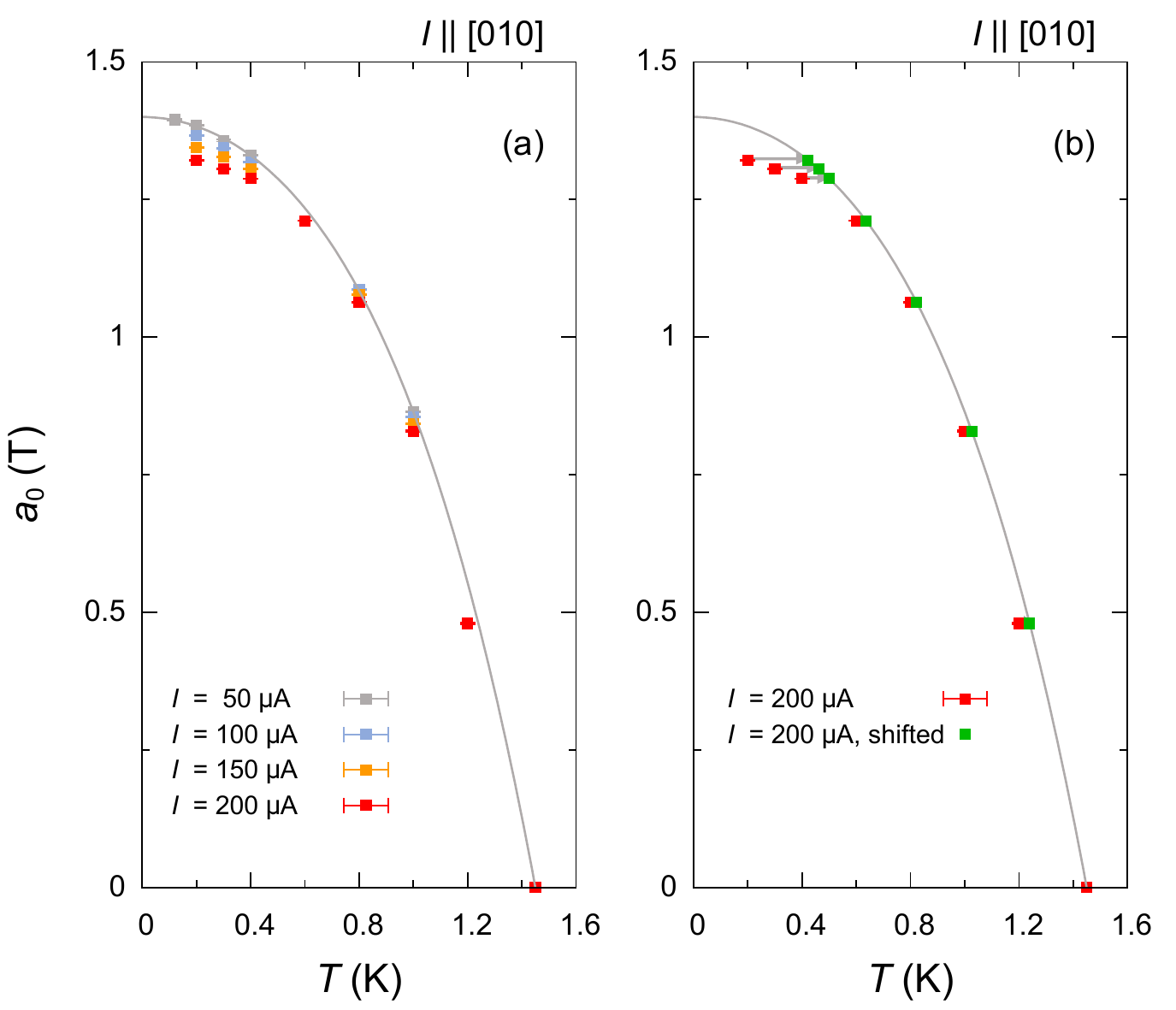}
      \caption{Examination of heating effect from the $T$-dependence of the coefficient $a_0$ considered as the in-plane average $H_{\soeji{c2}}$.
      (a) Gray, sky-blue, yellow and red squares show $a_0$ measured with current strength $I = 50, 100, 150$ and $200 \muup$A respectively.
      Gray curves shows the fitting curve obtained from the in-plane average $H_{\soeji{c2}}$ measured with 50$\muup$A   
      (b) Red squares and gray curves show the same data as those in the panel~(a).
      Green squares show the points with $a_0$ measured with $I = 200 \muup$A shifted onto the fitting curve.
      To estimate the deference between the actual sample temperature and the measured temperature,
        we assume that the change in $a_0$ is caused only by Joule heating and the gray curves
        exhibits the ideal $H_{\soeji{c2}}$-$T$ phase diagram of the sample.
      This estimation indicates the temperature increase as evaluated by the shift indicated by the gray arrows
       is at most 0.2~K in the lowest temperature region
       and less than 0.1~K above 0.4~K.
      Therefore, the current-heating effect has no qualitative effects on the analyses and conclusions.
      }\label{heating}
    \end{figure}

    \begin{figure}[htb]
      \centering
      \includegraphics[keepaspectratio, scale=0.90]{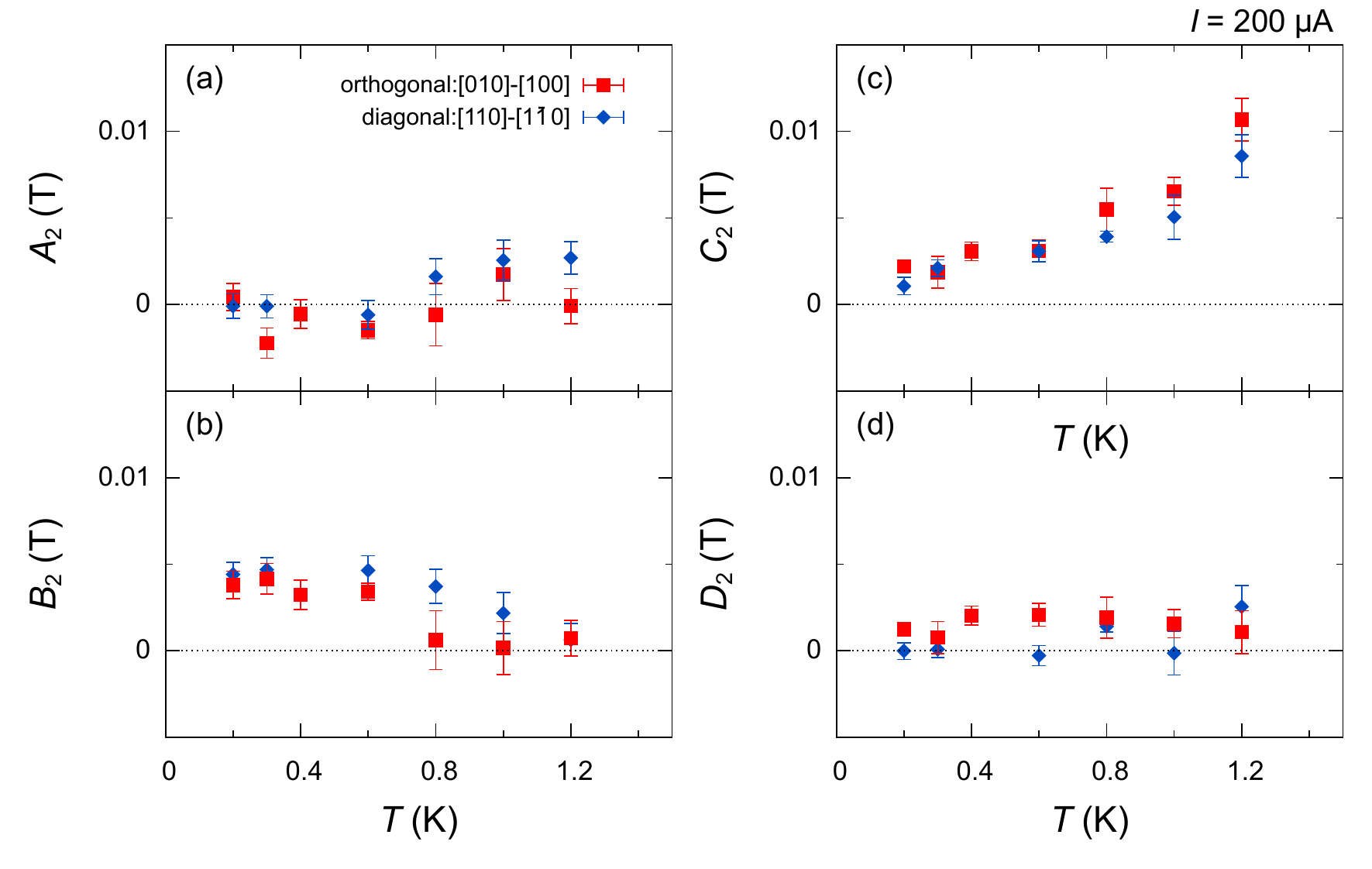}
         \caption{
        $T$-dependence of the two-fold $I$-dependent component and $I$-independent component
         of $H_{\soeji{c2}}^{(2)}$.
        Red squares and blue diamonds indicate values evaluated from the datasets for the orthogonal and diagonal
         current-directions, respectively.
         $B_2$ and $C_2$ shown in the panels (b) and (c) respectively are the same as $B_2$ and $C_2$ in Fig.~5 of the main text. 
        (a) Two-fold $I$-independent cosine component. This coefficient exhibits nearly zero at all temperatures.
        (b) Two-fold $I$-independent sine component. This coefficient shows about 0.005 T
         in the first-order transition region ($T\leq 0.8$~K) and nearly zero in the second-order transition region ($T\geq 0.8$~K).
        This fact implys that the sample in a high-field state gets sensitive to inhomogeneity effect producing two-fold anisotropy.
        (c) Coefficient for the $\cos 2(\phi_H -\phi_I)$ component, which increases on warming. It is consistent with vortex flow effect.
        (d) Coefficient for the $\sin 2(\phi_H -\phi_I)$ component, exhibiting nearly zero in the whole temperature range.  
          }
         \label{Capital2}
    \end{figure}



\end{document}